\documentclass[%
superscriptaddress,
groupedaddress,
unsortedaddress,
 amsmath,amssymb,
 aps,
twocolumn
]{revtex4-2}

\usepackage{graphicx}
\usepackage{dcolumn}
\usepackage{bm}
\usepackage{xcolor,soul}

\usepackage{hyperref}

\usepackage[normalem]{ulem}

\DeclareRobustCommand{\vect}[1]{
  \ifcat#1\relax
    \boldsymbol{#1}
  \else
    \mathbf{#1}
  \fi}

\newcommand{\cbr}[1]{\left(#1\right)}
\newcommand{\sbr}[1]{\left[#1\right]}

\newcommand{\tdep}[1]{_{#1}}
\newcommand{\idep}[1]{^{\scriptscriptstyle{(#1)}}}
\newcommand{\Thetadep}[1]{^{#1}}

\DeclareRobustCommand{\vc}[1]{
  \ifcat#1\relax
    \boldsymbol{#1}
  \else
    \mathbf{#1}
  \fi}
\newcommand{\mat}[1]{\mathsf{#1}}

\newcommand{\tzero}{t_0}
\newcommand{\tint}{t_1}
\newcommand{\ttwo}{t_2}

\newcommand{\q}{q\tdep{t}}
\newcommand{\p}{p\tdep{t}}
\newcommand{\U}{U\tdep{t}}
\newcommand{\w}{w\tdep{t}}

\newcommand{\ff}{f}
\newcommand{\vf}{v}
\newcommand{\uf}{u}

\newcommand{\sty}{y}
\newcommand{\cst}{w}

\newcommand{\Prob}{\mathcal{P}}
\newcommand{\pr}{f}
\newcommand{\dd}{\mathrm{d}}

\begin{document}

\title{Underdamped Stochastic Optimal Transport with Landauer boundary conditions: machine learning and analytical solution}

\author{Julia Sanders} 
\affiliation{Department of Mathematics and Statistics, University of Helsinki, Helsinki, 00014 Finland} 
\email{julia.sanders@helsinki.fi}

\author{Dario Lucente}
\affiliation{Department of Physics, Sapienza University of Rome, Rome, 00185, Italy}\email{dario.Lucente@roma1.infn.it}

\author{Marco Baldovin}
\affiliation{Institute for Complex Systems (CNR), Rome, 00185 Italy}
 \email{marco.baldovin@cnr.it}

\author{Paolo Muratore-Ginanneschi}
\affiliation{Department of Mathematics and Statistics, University of Helsinki, Helsinki, 00014 Finland} \email{paolo.muratore-ginanneschi@helsinki.fi}

\date{\today}

\begin{abstract}
Finding optimal transport protocols for underdamped stochastic processes is a difficult yet important problem in control theory. In this paper we develop a direct method, based on machine learning, to find the optimal solution of the problem in the presence of non-linear boundary conditions, including those modelling the Landauer bit-erasure problem. The dynamics is steered from a single-peak to a double-peak distribution in position space, by only acting on a controllable external potential. The problem is highly non-linear, and analytical solutions are not available in general. We benchmark our numerical results with a perturbative expansion, which is expected to hold in a suitable overdamped limit.
\end{abstract}

\maketitle


\section{Introduction}

Stochastic optimal transport addresses the problem of finding the (component of the) drift of a stochastic differential equation under the following two requirements: 
\begin{itemize}
    \item minimizing an assigned cost functional
    \item steering the evolution of an initially assigned probability distribution towards a target distribution
\end{itemize}

over a fixed time interval. Thus, stochastic optimal transport can be conceptualized as a generalization of the optimal mass transport problem~\ \cite{MonG1781,KanL1942,KanL1948}.
We refer to \cite{FigA2024} for an up-to-date concise review of the mathematical theory of optimal mass transport for deterministic flows, and to \cite{PeCu2019} for an overview of recent developments in computational methods. 

Stochastic mass transport was introduced in the physics literature by Erwin Schr\"odinger in \cite{SchE1931} (English translation in \cite{ChMGSc2021}). Schr\"odinger was interested in
characterizing the likelihood that a given thermodynamic state could be produced by a fluctuation from equilibrium.
Schr\"odinger's work~\cite{SchE1931} led to significant developments in the theory of stochastic processes~\cite{JamB1974,NelE2001}, and stochastic variational principles~\cite{FoeH1985,DaiP1991,CrZa1991}. A comprehensive survey of these mathematical developments is provided in \cite{LeoC2014}, and more recently in \cite{ChGePa2021}.

Over the last two decades, stochastic mass transport has become the focus of strong cross-disciplinary interest for independent but intertwined reasons. In control engineering, see e.g. \cite{TodE2009b}, identifying the cost of stochastic mass transport with the Kullback-Leibler divergence (e.g. \cite[\S~2.3]{CoTh2006}) between the realized and a reference process, typically free diffusion, can assess the performance of an actuator accomplishing an assigned task in the presence of noise. The stochastic optimal transport problem of minimizing the Kullback-Leibler divergence is usually referred to as the dynamic Schr\"odinger bridge \cite{LeoC2014}. In the overdamped formulation of the dynamic Schr\"odinger bridge, the optimal control problem is described by a Hamilton-Jacobi-Bellman equation that can be greatly simplified using the Cole-Hopf (or logarithmic) transform \cite[\S~VI]{FlSo2006}. This also is true in the underdamped case for controls that are functions of both position and momentum. Numerical solutions for large classes of boundary data can then be found using proximal algorithms \cite{CaHa2022} that exploit recent breakthroughs \cite{JoKiOt1998} in the theory of gradient flows, see \cite{SanF2017,AmBrSe2024} for thorough overviews.

In physics, interest in stochastic mass transport is motivated by advances in nano-system manipulation and the emergence of stochastic (or finite-time) thermodynamics \cite{PePi2020}. Motion at nano-scale is inherently affected by Brownian forces, whose effect cannot be avoided even
in a near vacuum \cite{KaLeZe2007}. A fundamental question is how nano-systems can efficiently harness randomness in order to move between assigned states or perform thermodynamic work on larger scales \cite{BrSi2019}. Considerable experimental efforts, see, e.g., \cite{BeArPeCiDiLu2012,KoMaPeAv2014}, focus on demonstrating optimal ``information ratchets'': control at minimum energy cost of the dynamics of a nano-system by changing the effective kinetic barriers to Brownian motion depending on the position of the system. Optimization of memory bits operations~\cite{lopez2016sub,dago2022,barros2025}, including Landauer bit-erasure~\cite{LeOrPoSn2019}, belongs to this class of problems. In stochastic systems, thermodynamic state indicators such as work, heat and entropy become fluctuating quantities whose statistics is governed by precise fluctuation relations, reviewed e.g. in \cite{JarC2011}. In particular, fluctuation relations express physical measures of dissipation in terms of information-theoretic measures of time-asymmetry, such as the Kullback-Leibler divergence between a process evolving forward-in-time and a backward-in-time counter-part \cite{ChGa2008}. This Kullback-Leibler divergence, identified as the mean entropy production, specifies a natural cost for stochastic optimal transport in finite-time thermodynamics: it vanishes if and only if the system is at thermal equilibrium. Interestingly, if  small-system dynamics is modelled by an overdamped (Langevin-Smoluchowski) stochastic differential equation (see e.g. \cite{PavG2014}), stochastic optimal transport admits an exact mapping into classical deterministic optimal mass transport \cite{AuGaMeMoMG2012}.

A second reason for interest for stochastic optimal transport, resulting in explosive growth in the number of publications, comes from generative modelling \cite{SoDiWeMaGa2015,HoAjAb2020,SoSoKiAbEr2021} in machine learning. A deep neural network can learn to generate data from noise by means of a two-step process. First, a forward stochastic process destroys structure in a data distribution by adding noise over a finite time interval. Next, a neural network is trained to revert the process by learning a backward evolution that interpolates between the initial data distribution and the one realized at the end of the forward evolution. Dynamic Schr\"odinger bridges provide a concrete implementation of this idea \cite{DeBoThHeAr2021,BlBeRiNe2025}, which 
we may also refer to as ``stochastic interpolant theory'' \cite{AlBoVaEI2023} and has a wide range of important applications, from sampling to image reconstruction \cite{YaSoHoXuZhYa2023}. 

These developments also have feedback to stochastic thermodynamics: they provide a framework of direct optimization methods \cite{DoEn2024} to numerically solve stochastic optimal transport problems, describing rectified motion or useful work extraction in a small system. A crucial difference from stochastic interpolant theory is that physically relevant models of nano-system dynamics evolve according to Langevin-Kramers dynamics. The Langevin-Kramers dynamics pose kinematic restrictions, including the coexistence of a conservative and a dissisipative component in the drift, and that control is exerted by varying a mechanical potential. How can computational strategies be adapted, and how can their accuracy be assessed in stochastic thermodynamics?

Specifically, we consider a controlled Langevin-Kramers dynamics. Control is exerted via a mechanical potential that, by definition, only depends upon the position coordinate of the physical system and not on its momentum. To avoid theoretical controllability issues, we do not directly consider a Schr\"odinger bridge problem but a weaker version, inspired by \cite{PaTaTr2021}. We add a terminal cost to the Kullback-Leibler divergence between the controlled process and a reference diffusion. This terminal cost is the static Kullback-Leibler divergence between the realized joint probability measure at the end of the control and a fixed, assigned measure. The terminal cost is weighed by a constant that we can increase to construct, as in augmented Lagrangian methods \cite{FoGl1983}, a family of control problems formally converging to a full dynamic Schr\"odinger bridge. 

There are two reasons motivating our choice of the running cost. First, as shown in \cite{SaBaMG2024}, this choice of the cost may be adapted to model an isochoric transition in a Brownian Carnot engine \cite{MaRoDiPePaRi2016}. Second, in contrast with the mean entropy production in the underdamped case, the running cost is always a quadratic functional of the control. This is a technical advantage for comparing our numerical machine learning results with analytic predictions obtained by multiscale perturbation theory around the overdamped limit \cite{PavG2014}. 

This comparison, together with the development of algorithmic methods for the kinematic constraint posed by the physically relevant formulation of the control problem, constitute the two main contributions of the present work.

\section{Formulation of the problem}
Consider the stochastic dynamics described by the Langevin equation
\begin{subequations}
\label{eq:dyn}
\begin{eqnarray}
    \dd\q&=&\varepsilon\, \p\, \dd t\\
    \dd\p&=&\cbr{-\p-\varepsilon \, \partial \U(\q) }\, \dd t + \sqrt{2}\,\dd\w\,,
\end{eqnarray} 
\end{subequations}
where $\q$ and $\p$ can be seen as the position and momentum, in dimensionless variables, of a colloidal particle subject to the external potential $\U(\q)$ and thermal noise, with a damping coefficient proportional to $\varepsilon^{-1}$. In the $\varepsilon \to 0$ limit, the dynamics tends to an overdamped regime. More details on the physical interpretation of the model can be found in~\cite{sanders2025minimal}.

The system is prepared at time $t=0$ according to an assigned probability density function (pdf) $\pr_0(q,p)=\rho_i(q,p)$. The Schr\"odinger bridge problem finds the time dependent protocol $\U(q)$ that, at time $t=T$, reaches a pre-determined final distribution $\rho_f(q,p)$, while minimizing the Kullback-Leibler divergence from a known reference dynamics
\begin{equation}
\begin{aligned}
\label{eq:kldyn}
&\mathrm{KL}^{\text{(dyn)}}\sbr{\Prob|\Prob^{(r)}}=\mathbb{E}_{\Prob} \!\sbr{\ln \cbr{\frac{\dd\Prob}{\dd\Prob^{(r)}}\cbr{\{\vc{x}\tdep{t}\}_{t\in[0,T]}}}}\\
&=\frac{\varepsilon^2}{4}\int_0^T \!\dd t\,\int_{\mathbb{R}^2}\! \dd q \, \dd p\, \pr_t(q,p) \cbr{\partial U\tdep{t}(q)-\partial U\idep{r}\tdep{t}(q)}^2.
\end{aligned}
\end{equation}
Here $\{\vc{x}\tdep{t}\}_{t\in[0,T]}$ denotes the path of the particle state $(\q,\p)$ in the time interval $[0,T]$, $\Prob$ represents the pdf of the path, while $\Prob^{(r)}$ is the pdf of the reference dynamics, generated by Eq.~\eqref{eq:dyn} with $\U\idep{r}(q)=U^{(r)}(q)$. In this paper, we fix the free diffusion, i.e. $\U\idep{r}(q)=0$ as the reference.

From a numerical point of view, imposing the constraint on the final pdf is very challenging. We therefore relax this constraint, instead imposing a terminal cost consisting of the static Kullback-Leibler divergence between the final distribution $\pr\tdep{T}(\vc{x})$ and the target one $\rho_f(\vc{x})$:
\begin{equation}
   \mathrm{KL}^{\text{(st)}}=\int_{\mathbb{R}^2} \dd \vc{x}\,\pr\tdep{T}(\vc{x})\ln \cbr{\frac{\pr\tdep{T}(\vc{x})}{\rho_f(\vc{x})}}\,.
\end{equation}
Therefore, the total cost to minimize is given by 
\begin{equation}\label{eq:action}
   \mathcal{C} =\mathrm{KL}^{\text{(dyn)}}\sbr{\Prob|\Prob^{(r)}}+M\, \mathrm{KL}^{\text{(st)}}\sbr{\pr\tdep{T}|\rho_f}\,.
\end{equation}

\subsection{Action for the variational problem}
The action in Pontryagin's formalism reads
\begin{equation}
\label{eq:action_pontryagin}
\begin{aligned}
    \mathcal{A}^{(M)}[\Prob,U,V]&=\mathrm{KL}^{\text{(dyn)}}\sbr{\Prob|\Prob^{(r)}}+M\, \mathrm{KL}^{\text{(st)}}\sbr{\pr\tdep{T}|\rho_f}\\
    &- \int_0^T \dd t \,\int_{\mathbb{R}^2}\dd \vc{x} \,V\tdep{t}(\vc{x})\cbr{\partial\tdep{t}-\mathcal{L}^{\dagger}\tdep{t}(\vc{x})}\pr\tdep{t}(\vc{x})\,,
\end{aligned}
\end{equation}
where $M$ is a positive constant,
\begin{equation}
\mathcal{L}^{\dagger}\tdep{t}=  -\varepsilon \cbr{p \partial_q -\cbr{\partial U\tdep{t}}\partial_p}+\partial_p \cbr{p+\partial_p}  
\end{equation}
is the forward Fokker-Planck operator that evolves $\pr\tdep{t}$ consistently with the Langevin dynamics~\eqref{eq:dyn}, and $V\tdep{t}$ is the so-called \textit{value function}, a time-dependent field that plays the role of a Lagrange multiplier, enforcing the dynamical evolution of $\pr\tdep{t}$. 
Since $M$ tunes the relative importance of the terminal cost, we expect the minimizer of $\mathcal{A}^{(M)}$ to tend to the solution of the Schr\"odinger bridge problem in the $M\to \infty$ limit.

\section{Analytical Methods}
The analytical solution of problem~\eqref{eq:action_pontryagin} relies on variational methods. The idea is detailed for instance in~\cite{SaBaMG2024}. At the optimal solution, the functional derivatives of the action with respect to  $V\tdep{t}$,  $\pr\tdep{t}$ and $\partial U\tdep{t}$ are set to zero. One gets:
\begin{subequations}
\label{eq:prob}
    \begin{align}
    \label{eq:fp}
        \partial_t \pr\tdep{t} -\mathcal{L}^{\dagger}\tdep{t}\pr\tdep{t}&=0\\
        \label{eq:vf}
        \partial_t V\tdep{t}+ \mathcal{L}\tdep{t} V\tdep{t}&=-\frac{\varepsilon^2}{4}\cbr{\partial U}^2\\
        \label{eq:sc}
        \int_{\mathbb{R}} dp \, \pr_t(q,p)\partial_p V\tdep{t}(q,p)&=\frac{\varepsilon\partial U\tdep{t}(q)}{2}   \int_{\mathbb{R}} dp\, \pr\tdep{t}(q,p)\,.\,\,    
    \end{align}
\end{subequations}
 Equation~\eqref{eq:fp} is just the Fokker-Planck dynamics of $\pr\tdep{t}$ (evolution of the state); Eq.~\eqref{eq:vf} is a differential equation that fixes the dynamics of $V\tdep{t}$ (evolution of the co-state, control equation); Eq.~\eqref{eq:sc} is an algebraic constraint (stationarity condition). The backward Fokker-Planck operator $\mathcal{L}\tdep{t}$ appearing in Eq.~\eqref{eq:vf} is defined as the adjoint of $\mathcal{L}^{\dagger}\tdep{t}$. The solution of the problem must satisfy the three equations~\eqref{eq:prob}, plus the boundary conditions given by the initial distribution and the terminal cost:
\begin{subequations}
\label{eq:bc}
\begin{align}    
\label{eq:bcf}
    \ff\tdep{0}(\vc{x})&=\rho_i(\vc{x})\,, \\
   \label{eq:bcv} V\tdep{T}(\vc{x})&=M\, \ln \cbr{\frac{\ff\tdep{T}(\vc{x})}{\rho_f(\vc{x})}}\,.
\end{align}
\end{subequations}

\subsection{Exact solution in the Gaussian case}
\label{sec:anlinear}
If $\rho_i$ and $\rho_f$ are Gaussian, the problem can be formulated as a boundary value problem for a system of ordinary differential equations for the distribution moments and coefficients of the control, and solved with standard numerical methods.  

Because a transition at minimum cost between two Gaussian states means that the distribution will remain Gaussian throughout the control horizon, we can make an ansatz that the gradient of potential will be linear
\begin{equation}
\begin{aligned}
    -\partial U\tdep{t}(q) = c\tdep{t} + k\tdep{t} q\,,
\end{aligned}
\end{equation}
where $c\tdep{t}$ and $k\tdep{t}$ are functions of time only, to be determined. In this case, the evolution of the probability distribution is fully specified by those of the first 
\begin{equation}
    \sty\idep{1}\tdep{t}=	\mathbb{E}[q\tdep{t}]\,,
		\quad \sty\idep{2}\tdep{t}=	\mathbb{E}[p\tdep{t}]\,
\end{equation}
and second order cumulants:
\begin{align}
	&
	\begin{split}
		&\sty\idep{3}\tdep{t}=\mathbb{E}\sbr{\left(q\tdep{t}-\sty\idep{1}\tdep{t}\right)^{2}}\,,
        \\
		&	\sty\idep{4}\tdep{t}=\mathbb{E}\sbr{\left(p\tdep{t}-\sty\idep{2}\tdep{t}\right)\left(q\tdep{t}-\sty\idep{1}\tdep{t}\right)}\,,
		\\
		&	\sty\idep{5}\tdep{t}=\mathbb{E}\sbr{\left(p\tdep{t}-\sty\idep{2}\tdep{t}\right)^{2}}\,.
	\end{split}
\end{align}

It is easy to verify that the vector $\vc{\sty}\tdep{t}$ of these quantities evolves according to
\begin{equation}
\label{eq:linfp}
\dot{\vc{\sty}}\tdep{t}=\begin{bmatrix}
    \mat{K}_I & 0\\
    0 & \mat{K}_{II}
\end{bmatrix}\, \vc{\sty}\tdep{t}+2 \vc{e}_5
\end{equation}    
 where the linear operators
  \begin{equation}  
    \quad \mat{K}_I=\begin{bmatrix}
        0   &   \varepsilon \\
        -\varepsilon k\tdep{t}   &  -1 
       \end{bmatrix},\quad \mat{K}_{II}=
       \begin{bmatrix}
           0   &   2\varepsilon   &   0 \\
           -\varepsilon k\tdep{t}   &   -1   &  \varepsilon \\
           0   &   -2\varepsilon k\tdep{t}   &   -2  
    \end{bmatrix}\,
\end{equation}
act on the subspace of the first and second cumulants, respectively, and $\vc{e}_n$ is a vector such that $e_n^{(j)}=\delta_{nj}$. Equation~\eqref{eq:linfp} determines the evolution of the state. Recalling that any bi-variate Gaussian distribution can be written as
\begin{equation}
    \pr\tdep{t}(q,p)=\frac{\exp \cbr{-(\vc{x}-\mathbb{E}[\vc{x}\tdep{t}])^{\top} \mat{C}\tdep{t}^{-1} (\vc{x}-\mathbb{E}[\vc{x}\tdep{t}])}}{2 \pi \sqrt{|\det \mat{C}\tdep{t}|}}
\end{equation}
with 
$$
\mathbb{E}[\vc{x}\tdep{t}]=\begin{bmatrix} \sty\tdep{t}\idep{1}\\ \sty\tdep{t}\idep{2}    
\end{bmatrix}\,, \quad \mat{C}\tdep{t}=\begin{bmatrix}
   \sty\idep{3}\tdep{t}&\sty\idep{4}\tdep{t}\\ 
    \sty\idep{4}\tdep{t}&\sty\idep{5}\tdep{t}\\ 
\end{bmatrix}\,,
$$
and introducing the explicit expression
\begin{equation}
    V\tdep{t}(q,p)=\cst\idep{0}\tdep{t}+\cst\idep{1}\tdep{t}q+\cst\idep{2}\tdep{t}p+\cst\idep{3}\tdep{t}q^2+\cst\idep{4}\tdep{t}qp+\cst\idep{5}\tdep{t}p^2\,,
\end{equation}
one finds that the evolution of the co-state reads
\begin{equation}
\label{eq:lince}
    \dot{\vc{\cst}}\tdep{t}=-\begin{bmatrix}
    \mat{K}_I^{\top} & 0\\
    0 & \mat{K}_{II}^{\top}
\end{bmatrix}\, \vc{\cst}\tdep{t}-\cbr{\frac{\varepsilon k\tdep{t}}{2}}^2 \vc{e}_3\,.
\end{equation}
Finally, the stationary condition imposes
\begin{subequations}
\label{eq:linsc}
\begin{eqnarray}
c\tdep{t}&=&\frac{2 \cst\idep{2}}{\varepsilon}-k\tdep{t} \sty\idep{1}\\
k\tdep{t}&=&2\,\frac{\sty\tdep{t}\idep{3}\cst\tdep{t}\idep{4}+2\,\sty\tdep{t}\idep{4}\cst\tdep{t}\idep{5}}{\varepsilon\,\sty\tdep{t}\idep{3}}\,.    
\end{eqnarray}
\end{subequations}
Once inserted into Eqs.~\eqref{eq:linfp},~\eqref{eq:lince}, this relation leads to a system of 10 scalar, nonlinear ODEs for $\vc{\sty}\tdep{t}$ and $\vc{\cst}\tdep{t}$. Together with the boundary conditions~\eqref{eq:bc}, these result in boundary value problem that can be solved via standard numerical methods, such as collocation or shooting~\cite{RaoA2010}.

\subsection{Multiscale perturbation theory}
\label{sec:multiscale}
In the case of non-Gaussian initial or target pdf, exact analytical methods to the solution of our problem are unknown. If the dynamics is closed to the overdamped limit (i.e., if $\varepsilon \ll 1$), we can however rely on a perturbative strategy,  which has been already applied to the ``full-bridge'' case in a previous work by some of us~\cite{SaBaMG2024}. The calculation is long, and very close to the one presented in~\cite{SaBaMG2024}, so, we only report the main steps of the procedure, and highlight any differences with the ``full-bridge'' case. 

\subsubsection{Hermite polynomial expansion}
We start by expanding the unknowns of problem~\eqref{eq:prob} in Hermite polynomials of the momentum. We recall that those are defined as
$$
\begin{aligned}
    H_{0}(p)&=1\,,\\
\partial_p H_{n}(p)&=nH_{n-1}(p)    \quad n\ge1\
\end{aligned}
$$
and satisfy the orthogonality condition
$$
\int_{\mathbb{R}}dp\, \frac{e^{-p^2/2}}{\sqrt{2 \pi}}H_{n}(p) H_{m}(p) = n!\, \delta_{n m}\,.
$$
We introduce the notation
\begin{subequations}
\label{eq:herexp}
\begin{align}
    \pr\tdep{t}(q,p)&=\frac{e^{-p^2/2}}{\sqrt{2 \pi}} \sum_{n=0}^{\infty} \ff\idep{n}\tdep{t}(q)H_n(p)\\
    V\tdep{t}(q,p)&= \sum_{n=0}^{\infty} \vf\idep{n}\tdep{t}(q)H_n(p)\,, 
\end{align}
\end{subequations}
where $\{\ff\idep{n}\tdep{t}\}_{n\in\mathbb{N}}$ and $\{\vf\idep{n}\tdep{t}\}_{n\in\mathbb{N}}$ are functions of the position $q$ that need to be determined. By plugging Eqs.~\eqref{eq:herexp} into~\eqref{eq:prob} and projecting onto $H_n(p)$, we get an infinite hierarchy of PDEs in $q$ and $t$ (instead of two PDEs in $q$, $p$ and $t$), plus the stationarity condition~\eqref{eq:sc} and the boundary conditions~\eqref{eq:bc}.

\subsubsection{Perturbative \texorpdfstring{$\varepsilon$}{Ɛ}-expansion}
We are now in a position to address the problem via a perturbative multiscale scheme. The idea is to expand  each of the functions $\{\ff\idep{n}\tdep{t}\}_{n\in\mathbb{N}}$ and $\{\vf\idep{n}\tdep{t}\}_{n\in\mathbb{N}}$, and the drift $\partial U$, as power series of $\varepsilon$. At the same time, we allow each coefficient of such expansion to depend on two time scales, namely on the auxiliary time variables
$$
\tzero =t\,,\quad \ttwo=\varepsilon^2 t.
$$
The dependence on these artificial time scales introduces a larger number of tunable parameters with respect to ordinary perturbative schemes. This additional freedom can be used to systematically cancel secular terms that are known to arise, for instance, when dealing with problems characterized by boundary layers~\cite{PaSt2008}. For a generic time-dependent function $h\tdep{t}$, one has therefore
\begin{equation}
\label{eq:diffrule}
    \partial_t h\tdep{t} =   \partial_{\tzero} h\tdep{\tzero,\ttwo} + \varepsilon^2 \partial_{\ttwo}h\tdep{\tzero,\ttwo}\,.
\end{equation}
The unknowns can be thus expanded as:
\begin{subequations}
\label{eq:epsexp}
    \begin{align}
       \ff\idep{n}\tdep{t}(q)&=\sum_{j=0}^{\infty}\varepsilon^j \ff\idep{n:j}\tdep{\tzero,\ttwo}(q)\,,\\
   \vf\idep{n}\tdep{t}(q)&= \sum_{j=0}^{\infty} \varepsilon^j\vf\idep{n:j} \tdep{\tzero,\ttwo}(q)\,,\\
    U\tdep{t}(q)&= \sum_{j=0}^{\infty} \varepsilon^j\uf\idep{j} \tdep{\tzero,\ttwo}(q)\,.
    \end{align}
\end{subequations}
After inserting these expressions into the ODE hierarchy and applying the differentiation rule~\eqref{eq:diffrule}, we can isolate the coefficients of the same power of $\varepsilon$.

\subsubsection{Boundary conditions}
Before starting the iterative solution of the perturbative problem, one also needs to determine the boundary conditions for the new set of variables. With slight abuse of notation, we adopt the convention
$$
\begin{aligned}
    h\tdep{0}&=h\tdep{0,0}\\
    h\tdep{T}&=h\tdep{T,\varepsilon^2 T}\,.    
\end{aligned}
$$
We assume that the initial and the target pdf have the Botzmann-Gibbs form
$$
\rho_{i,f}(q,p)=\frac{e^{-p^2/2-U_{i,f}(q)}}{\sqrt{2 \pi}\int_{\mathbb{R}}\dd q\, e^{-U_{i,f}(q)} }
$$
where $U_i(q)$ and $U_f(q)$ are some confining potentials. By defining the marginal pdfs
$$
\widetilde{\rho}_{i,f}(q)=\int_{\mathbb{R}}dp\, \rho_{i,f}(q,p)\,,
$$
 we notice that the boundary condition~\eqref{eq:bcf} is compatible with
\begin{equation}
\label{eq:bcf1}
        \ff\idep{n:i}\tdep{0}(q)= \widetilde{\rho}_{i}(q)\, \delta_{n\,0}\delta_{i\,0}\,
\end{equation}
for the pdf. Of course, other choices would have been equally valid, provided that Eq.~\eqref{eq:bcf} is fulfilled. We exploit a similar freedom at the final time $t=T$ to impose the condition
$$
\frac{\ff\idep{n}\tdep{T}(q)}{\ff\idep{0}\tdep{T}(q)}H_n(p)=O(\varepsilon^n)\,,
$$
which in turns allows us to write Eq.~\eqref{eq:bcv} as
\begin{subequations}
\label{eq:bcv1}
    \begin{align}
    \label{eq:bcv10}
        \vf\idep{0:j}\tdep{T}&=M \ln \cbr{\frac{\ff\idep{0:0}\tdep{T}(q)}{\widetilde{\rho}_{f}(q)}}\, \delta_{j 0}\\
        \nonumber &+ \dfrac{M}{2} \cbr{2\, \frac{    \ff\idep{0:2}\tdep{T}(q) }{\ff\idep{0:0}\tdep{T}(q)} -\cbr{\frac{\ff\idep{1:1}\tdep{T}(q)}{\ff\idep{0:0}\tdep{T}(q)}}^2}\, \delta_{j 2}\,,\\
        \label{eq:bcv11}
        \vf\idep{1:j}\tdep{T}&= M \frac{\ff\idep{1:1}\tdep{T}(q)}{\ff\idep{0:0}\tdep{T}(q)}\, \delta_{j 1}\,,\\
        \label{eq:bcv12}
        \vf\idep{2:j}\tdep{T}&= \dfrac{M}{2} \cbr{2\, \frac{    \ff\idep{2:2}\tdep{T}(q) }{\ff\idep{0:0}\tdep{T}(q)} -\cbr{\frac{\ff\idep{1:1}\tdep{T}(q)}{\ff\idep{0:0}\tdep{T}(q)}}^2}\, \delta_{j 2}\,.
    \end{align}
\end{subequations}
The equations above are consistent with the request that, in the $M\to\infty$ limit, the problem tends to a full bridge and this limit is considered before the limit $\varepsilon\to 0$. Indeed, in order for the value function to stay finite, $\ff\idep{0:0}\tdep{T}(q)$ must tend to the position marginal pdf in the limit, while the higher order correction terms $\ff\idep{1:1}\tdep{T}(q)$ and $\ff\idep{2:2}\tdep{T}(q)$  of the pdf at the final time must be negligible with respect to $\ff\idep{0:0}\tdep{T}(q)$.

Terms corresponding to higher order Hermite polynomials can be also computed in principle, but we stop at second order. We emphasize that the only difference with respect to the case studied in~\cite{SaBaMG2024} are the boundary conditions~\eqref{eq:bcv1}.

\subsubsection{Iterative solution of the multiscale}
\begin{figure}
    \centering
    \includegraphics[width=.9\linewidth]{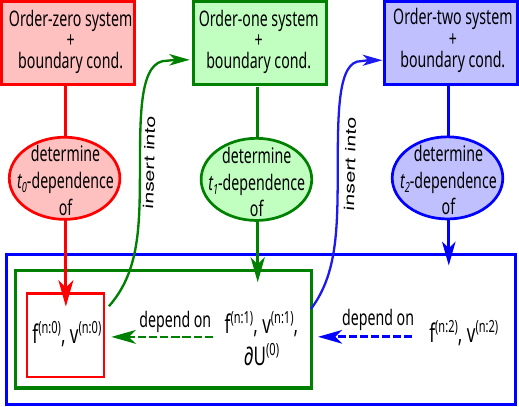}
    \caption{Visual summary of the multiscale perturbative scheme adopted to solve the problem in the $\varepsilon \ll 1$ limit.}
    \label{fig:scheme}
\end{figure}

At this point, we are in the position to perform an order-by-order study of our hierarchy. The strategy is summarized in Fig.~\ref{fig:scheme}. We start by solving the zeroth order system, complemented by the boundary conditions~\eqref{eq:bcf1} and ~\eqref{eq:bcv10}. This gives an analytical expression for the $\tzero$ dependence of  $\{\ff\idep{n:0}\tdep{\tzero, \ttwo}\}_n$ and $\{\vf\idep{n:0}\tdep{\tzero, \ttwo}\}_n$, leaving the dependence on the slower time scale $\ttwo$ still undetermined. These solutions can be substituted into the $O(\varepsilon)$ system, which can in turn be solved together with the boundary conditions~\eqref{eq:bcf1} and~\eqref{eq:bcv11}. By doing this, we determine expressions for  $\{\ff\idep{n:1}\tdep{\tzero, \ttwo}\}_n$, $\{\vf\idep{n:1}\tdep{\tzero, \ttwo}\}_n$ and $\{\uf\idep{0}\tdep{\tzero, \ttwo}\}_n$ in terms of $\{\ff\idep{n:0}\tdep{\tzero, \ttwo}\}_n$ and $\{\vf\idep{n:0}\tdep{\tzero, \ttwo}\}_n$. Moreover, 
one shows that there is no dependence on the timescale $\tint=\varepsilon t$ of any of the above mentioned variables. Finally, as a last step, we repeat the operation at order $O(\varepsilon^2)$, taking again into account Eqs.~\eqref{eq:bcf1} and~\eqref{eq:bcv12}, thus obtaining also the $t_2$ dependence of all the previous functions, as well as those of $\{\ff\idep{n:2}\tdep{\tzero, \ttwo}\}_n$ and $\{\vf\idep{n:2}\tdep{\tzero, \ttwo}\}_n$.
We report here only the main results of this order-by-order iterative procedure, referring again the interested reader to the supplementary material~\cite{SM} for detailed calculations.

At order $O(1)$ one straightforwardly gets: 
\begin{subequations}
    \begin{align}
        \ff\idep{n:0}\tdep{\tzero, \ttwo}(q)&=\delta_{n 0}\ff\idep{0:0}\tdep{0, \ttwo}(q)\,,\\ 
        \vf\idep{n:0}\tdep{\tzero, \ttwo}(q)&=\delta_{n 0}M \ln \cbr{\frac{\ff\idep{0:0}\tdep{T}(q)}{\widetilde{\rho}_f(q)}}+(1-\delta_{n0})\vf\idep{n:0}\tdep{T, \ttwo}(q)e^{-n(T-\tzero)}  \,.
    \end{align}
\end{subequations}
We determine the functions $\vf\idep{n:0}\tdep{T, \ttwo}(q)$  by consistency with the terminal cost.
When examining the system at $O(\varepsilon)$, one finds a relation for the leading term of the control:
\begin{equation}
   \partial \uf\idep{0}\tdep{\tzero, \ttwo}=-\frac{ \partial_{t_0}\ff\idep{1:1}\tdep{\tzero, \ttwo}+\ff\idep{1:1}\tdep{\tzero, \ttwo}+\partial_q \ff\idep{0:0}\tdep{0,\ttwo}}{\ff\idep{0:0}\tdep{0,\ttwo}}\,.
\end{equation}
The perturbative pdf correction $\ff\idep{1:1}\tdep{\tzero, \ttwo}$ is obtained by solving the ODE
\begin{equation}
\begin{aligned}
\label{eq:odef11}
    \partial\tdep{\tzero}^2 \ff\idep{1:1}\tdep{\tzero,\ttwo}-\ff\idep{1:1}\tdep{\tzero,\ttwo}=&\partial_q \ff\idep{0:0}\tdep{0,\ttwo}+2 \ff\idep{0:0}\tdep{0,\ttwo} \partial_q \vf\idep{0:0}\tdep{T,\ttwo} \\
    &+ 4 e^{-2(T-\tzero)}\partial_q\cbr{\vf\idep{2:0}\tdep{T,\ttwo}\ff\idep{0:0}\tdep{0,\ttwo}}\,,
    \end{aligned}
\end{equation}
with the boundary conditions
\begin{subequations}
    \begin{align}
        \ff\idep{1:1}\tdep{0,\ttwo}&=0\,,\\
        \ff\idep{1:1}\tdep{T,\ttwo}&=-\frac{\partial_q \ff\idep{0:0}\tdep{0,\ttwo} + \partial\tdep{\tzero} \widetilde{\ff}\idep{1:1}\tdep{\tzero,\ttwo}\Big|\tdep{\tzero=T} }{\coth T+1+2M+4 \vf\idep{2:0}\tdep{T,\ttwo}}\,.
    \end{align}
\end{subequations}
In the previous equation, $\widetilde{\ff}\idep{1:1}\tdep{\tzero,\ttwo}$ represents the solution of the ODE~\eqref{eq:odef11} in the full-bridge case, i.e. with boundary conditions
$$
\ff\idep{1:1}\tdep{0,\ttwo}=\ff\idep{1:1}\tdep{T,\ttwo}=0\,.
$$

Finally, at order $O(\varepsilon^2)$ the problem is closed by solving the cell problem
\begin{subequations}
    \begin{align}
        \partial\tdep{\ttwo}\ff\idep{0:0}\tdep{\tzero,\ttwo}-\partial_q \cbr{\ff\idep{0:0}\tdep{\tzero,\ttwo}\partial_q \sigma\tdep{\ttwo}}-\sqrt{A}\partial_q^2\ff\idep{0:0}\tdep{\tzero,\ttwo}&=0\\
        \partial\tdep{\ttwo}\sigma\tdep{\ttwo}-\frac{1}{2}\cbr{\partial_q \sigma\tdep{\ttwo}}^2+\sqrt{A} \partial_q^2 \sigma\tdep{\ttwo} &=0\,,
    \end{align}
\end{subequations}
where  $ \sigma\tdep{\ttwo}$ is an auxiliary variable and
$$
A=1-3\frac{\cbr{1+e^{-T}}^2}{1+4e^{-T}+e^{-2T} }\frac{\tanh(T/2)}{T}\,.
$$
The boundary conditions are specified in~\cite{SM}.

Using the above results, one can write explicit expressions for the corrections of the pdf and the control up to order $O(\varepsilon^2)$. They are involved and not very insightful: the interested reader can find them in~\cite{SM}.

\section{Numerical methods}
\label{sec:numerical}
In this section, we discuss the numerical method for minimizing the generalized underdamped Schrödinger bridge. For this, we take a direct approach and minimize instead on the action~\eqref{eq:action} over sampled trajectories~\cite{QuBo2024,BoQuRiSc2024,ExTh2018}. Unlike the analytic method, we do not pass through the first order optimality conditions.

The idea is as follows. We minimize the loss function given by~\eqref{eq:action}, where the control is parametrized by a neural network. We draw inspiration from~\cite{HuLaVa2024} to formulate the minimization in a computationally friendly way using Girsanov theorem. With the addition of a weighting constant $M>0$ in front of the terminal cost, we can then push the system towards the target final distribution, which, for sufficiently large $M$, means that the system reaches the target final boundary condition and we approximate the solution to the full bridge problem. 

We denote a neural network with a set of parameters $\Theta$ as $\mathbb{U}\Thetadep{\Theta}$. We use a feed-forward neural network, meaning a composition of affine (known as layers) and non-linear (known as activation) transforms. The vector $\Theta$ denotes all the parameters associated with affine transformations. Details of the parametrisation are described in~\cite{SM}.  

The dynamics' control is mapped into the network under the transformation 
\begin{align}
    \label{nn_to_du}
    \partial_q U_t(q) &= -\sqrt{2}\,\mathbb{U}\Thetadep{\Theta}(t,q)\,.
\end{align}

To find the optimal solution for the bridge, we need to find the set of parameters $\Theta$ that minimize the loss function~\eqref{eq:action}. This is done by a gradient descent, which means small, iterative updates to $\Theta$ in the direction that minimizes gradient of the loss with respect to $\Theta$. 

In practice, to compute the gradient of~\eqref{eq:action} with respect to $\Theta$, we need to use simulated trajectories of the dynamics~\eqref{eq:dyn}. Trajectories are simulated with a discretisation scheme over a time-grid, $0=t_0<t_1<\ldots<t_N=T_f$, with steps denoted by $\Delta t_i = t_{i+1}-t_i$. The distribution achieved at final time is estimated numerically from evolved samples, denoted by $\hat{f}\Thetadep{\Theta}$, where $\Theta$ corresponds to the parameters of the neural network defining the control protocol~\eqref{nn_to_du} used to obtain the samples. Therefore, the loss~\eqref{eq:action} of the numerical problem becomes
\begin{equation}\label{eq:discrete_loss}
    \mathcal{C}(\Theta) = \mathbb{E}_{\!\mathcal{P}\Thetadep{\Theta}}\biggl[\sum_{i=0}^{N-1} \dfrac{\Delta t_i}{2}\|\mathbb{U}_i\Thetadep{\Theta}\|^2  + M\,\ln\dfrac{\hat{f}\Thetadep{\Theta}\tdep{T}(q_T,p_T)}{\rho_f(q\tdep{T},p\tdep{T})}\biggr]\,,
\end{equation}
where we used the shorthand notation $\mathbb{U}_i\Thetadep{\Theta}=\mathbb{U}\Thetadep{\Theta}(t_i,q\tdep{t_i})$.
 The expectation is taken over the path measure for trajectories where the control is parametrized by $\Theta$, which is denoted by $\mathcal{P}\Thetadep{\Theta}$. The computation of the gradient of~\eqref{eq:discrete_loss} with respect to $\Theta$ is costly, since the simulated positions and momenta $p_t,q_t$ have an implicit dependence on the neural network parameters. The gradient calculation would involve the entire simulated trajectory. 

\subsection{Girsanov Formula}
We leverage Girsanov theorem \cite{GirI1960}, see e.g. \cite{KleF2005} for a modern presentation with applications, to simulate trajectories of the dynamics independently of the parameters to be updated in the neural network. This means that the positions and momenta values in~\eqref{eq:action} are independent of $\Theta$. To this end, we begin by making a copy of the parameters of the neural network
\begin{equation}
\label{theta_to_thetabar}
    \bar{\Theta} =\textbf{stopgrad}(\Theta)\,.
\end{equation}
This copy of the parameters is fixed throughout each iteration. 

The underdamped dynamics are degenerate, meaning that only the velocity $p_t$ has random noise and its diffusion matrix is not invertible. This means that we can only change the potential term and must leave the position drift in tact to ensure that we stay in the invertible part of the diffusion matrix. The change in the drift at time $t_i$ is given by 
\begin{align*}
    b\begin{pmatrix}
        p_{t_i} \\
        q_{t_i}
    \end{pmatrix} &= \begin{pmatrix}0 \\
        \sqrt{2}\mathbb{U}\Thetadep{\bar{\Theta}}_i-\sqrt{2}\mathbb{U}\Thetadep{\Theta}_i
    \end{pmatrix}\,,
\end{align*}
giving the resulting Girsanov weighting factor
\begin{align}
    \label{eq:rn_deriv}\begin{split}
    \mathcal{G}_{t_j}\! = \!\exp\biggl(&\!\dfrac{1}{4}\!\sum_{i=0}^{j-1}\!\Delta t_i \, |\mathbb{U}\Thetadep{\bar{\Theta}}_i\! -\!\mathbb{U}\Thetadep{\Theta}_i|^2 \!+\! \dfrac{1}{\sqrt{2}}\!\sum_{i=0}^{j-1} \langle\, \Delta w_i ,\, \mathbb{U}\Thetadep{\bar{\Theta}}_i -\mathbb{U}\Thetadep{\Theta}_i \,\rangle\!\biggr)
    \end{split}
\end{align}
with $\Delta w_i = w\tdep{t_{i+1}}-w\tdep{t_i}$ denoting a step of the Weiner process in the numerical discretisation scheme used for integrating the dynamics~\eqref{eq:dyn}. 

By Girsanov theorem, multiplying by the Girsanov weighting factor in the expectation~\eqref{eq:discrete_loss}, we can instead evaluate the expectation on trajectories where the protocol is parametrized with $\bar{\Theta}$. This path measure is denoted by $\mathcal{P}\Thetadep{\bar{\Theta}}$, and means that~\eqref{eq:discrete_loss} can be written as 
\begin{equation}
\label{eq:discrete_loss_bar}
    \mathcal{C}(\Theta)
    \!=\! \mathbb{E}_{\!\mathcal{P}\Thetadep{\bar{\Theta}}}\!\Biggl[\!\mathcal{G}_{t_N}\!\times\! \biggl(\sum_{i=0}^{N-1} \!\frac{\Delta t_i}{2}\|\mathbb{U}_{i}\Thetadep{\Theta}\|^2   + M \ln\!\dfrac{\hat{f}\Thetadep{\bar{\Theta}}\tdep{T}(q_T,p_T)}{\rho_f(q\tdep{T},p\tdep{T})}\!\biggr)\!\Biggr]\,.
\end{equation}
The gradient of~\eqref{eq:discrete_loss_bar} can now be found by a direct computation, since we no longer need to differentiate each position and momentum value. Finally, by direct calculation following~\cite{HuLaVa2024}, we notice that the alternative loss function 

\begin{equation}\label{eq:discrete_loss2}
    \begin{split}\Tilde{\mathcal{C}}(\Theta)
    = \mathbb{E}_{\mathcal{P}\Thetadep{\bar{\Theta}}}\!\Biggl[\!A(\Theta) +B(\Theta)\biggl(\!A(\bar\Theta) + M\ln\dfrac{\hat{f}\Thetadep{\bar{\Theta}}\tdep{T}(q_T,p_T)}{\rho_f(q\tdep{T},p\tdep{T})}\!\biggr)\!\Biggr]
\end{split}\end{equation}
with
$$
A(\Theta) = \sum_{i=0}^{N-1}\! \dfrac{\Delta t_i} {2}\|\mathbb{U}_i\Thetadep{\Theta}\|^2,\qquad B(\Theta) =\sum_{i=0}^{N-1} \!\langle \Delta w_i,\mathbb{U}_i\Thetadep{\Theta}\rangle
$$
has the same gradient as the loss function~\eqref{eq:discrete_loss_bar}. The loss~\eqref{eq:discrete_loss2} is easier to evaluate numerically, since the dependence on $\Theta$ only presents as sums.

\subsection{Final state estimation}
In order to fully evaluate the loss function~\eqref{eq:discrete_loss2}, we need to evaluate the final achieved distribution for any control $\mathbb{U}\Thetadep{\bar{\Theta}}$. Allowing a free form on the control means that there is no closed form of the final distribution. We can however generate samples from this distribution by evaluating the forward dynamics~\eqref{eq:dyn}. There are many numerical inference methods to estimate a distribution from samples. The challenge is to use minimum number of samples possible, as generating these samples requires forward evaluations of the neural network $\mathbb{U}\Thetadep{\bar{\Theta}}$.

From a mathematical and physical point of view, it makes sense to exploit the physical properties of the problem. For example, when the initial assigned distribution and target distribution are both Gaussians, we know that the optimal solution will ensure that the intermediate distributions are also Gaussian. We can therefore use the knowledge that the final achieved distribution will also be Gaussian, and use a parametric method, namely assume the Gaussian form and use the sample mean and variance as estimates. This reduces the last term of~\eqref{eq:action} to a closed functional form that can be evaluated.  

We also note that the final distribution does not need to be normalized, giving us a further computational saving. This is because of the form of the loss~\eqref{eq:discrete_loss2}, and the logarithm in front of the distributions. The normalisation constants can be extracted as constant functions of the position and momentum, and then multiplied by the Ito integral over the Weiner process. This expectation evaluates to zero.   

In general however, we do not have any information about the functional form of the final distribution, but, we do know that the final distribution is the solution at final time to the Fokker-Planck equation~\eqref{eq:fp}. We can therefore can instead use a numerical method specifically for the Fokker-Planck equation. Since we can evaluate trajectories of the dynamics~\eqref{eq:dyn} relatively straightforwardly, we use a Monte-Carlo based method which is valid for any mechanical, confining potential~\cite{SaMG2025}. This is derived with a change of variables in the underlying Fokker-Planck equation and a Girsanov factor applied to averages of integrals of backward trajectories of the dynamics~\eqref{eq:dyn}. The probabilities are computed only where they are needed, i.e. for those coordinates in the training set, as a numerical average of simulated trajectories that were started at those points. We refer to Section 3 of~\cite{SaMG2025} for details of the method.

\subsection{Summary of Algorithm}
In this section, we present a general summary of the machine learning method. For full technical specification of the training loop including optimizer and learning rate, technical details of the architecture of the neural network used and full code for the examples shown can be found in the supplementary material~\cite{SM}.

To find the optimal parameters $\Theta$ for the network $\mathbb{U}\Thetadep{\Theta}$ that minimize~\eqref{eq:discrete_loss2}, we use a gradient descent, known as the training process. Firstly, the dynamics~\eqref{eq:dyn} with the parametrisation~\eqref{nn_to_du} and where the parameters of the protocol are given by $\bar{\Theta}$~\eqref{theta_to_thetabar} are evolved from time $t=0$ under a suitable discretisation scheme. This means In practice, we may a leapfrog verlet scheme or a BAOAB scheme, see supplementary material~\cite{SM}. Along trajectories, we evaluate and store the summands $\Delta t_i \|\mathbb{U}\Thetadep{\Theta}\!(t_i,q\tdep{t_i})\|^2$ and $\langle\Delta w_i  \,,\,\mathbb{U}\Thetadep{\Theta}\!(t_i,q\tdep{t_i})\rangle$. 

At final time $T$, we have a set of final positions and momenta $(q_T,p_T)$. We initialize a number of trajectories starting at these coordinates that we then use to evaluate $\hat{f}\Thetadep{\bar{\Theta}}$ according to~\cite{SaMG2025}. We can then put together the loss function~\eqref{eq:discrete_loss2}, and perform a pass of backpropagation to evaluate the gradient. This is done with a machine learning library, Pytorch. 
The computed gradient is used to update the parameters $\Theta$.
\begin{align*}
    \Theta = \Theta - \eta\, \nabla_\Theta \,\mathcal{C}(\Theta)
\end{align*}
with $\eta>0$ known as a learning rate. In practice, the behavior of the gradient update step and value of $\eta$ is determined by an optimizer and learning rate scheduler, as it can depend on the size of the gradient. The behavior gradient update step is tailored to the problem, see supplementary material~\cite{SM}. After this step, the fixed parameters~$\bar{\Theta}$ are updated, so that~\eqref{theta_to_thetabar} is true.

We recall that the loss function used~\eqref{eq:discrete_loss2} is not convex in the control. Therefore we regularly evaluate the true loss function~\eqref{eq:action} to estimate convergence. We evaluate this using simulated trajectories of the dynamics under $\mathbb{U}\Thetadep{\bar{\Theta}}$ and the reference process, a free diffusion. To estimate the static Kullback-Leibler divergence between the achieved and target distribution, we use the final positions and momenta of the sample trajectories and samples from the target, and use the relative differences between their nearest neighbors, as described in~\cite{WaKuVe2009}. Gradient descent is performed for a fixed number of epochs, until there is no continuous improvement in numerical evaluations of~\eqref{eq:action}.  

\section{Results}
In this Section we show the solution to the optimal problem for different choices of the boundary conditions, of the damping parameter $\varepsilon^{-1}$ and of the constant $M$, specifying the relative weight between running and terminal cost in Eq.~\eqref{eq:action}.

\begin{figure}
    \centering
    \includegraphics[width=.99\linewidth]{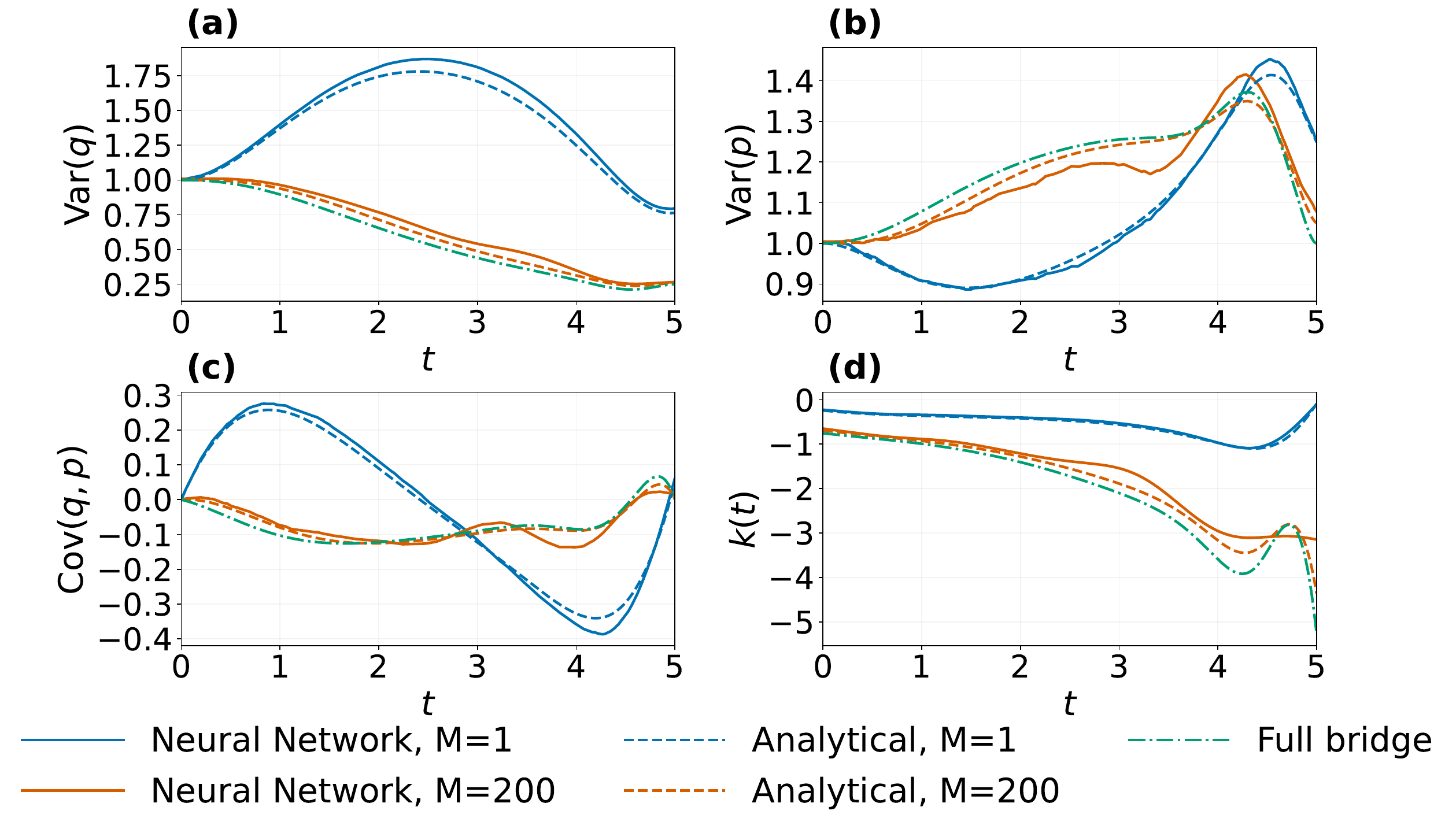}
    \caption{Solution of the optimal problem with Gaussian boundary conditions: $U_i(q)=q^2/2$, $U_f(q)=2q^2$. Panels (a) - (c) show second order cumulants, while the parameter $k(t)$ of the linear drift [see Eq.~\eqref{eq:linsc}] is plotted in (d). Solid lines represent the  neural network solution, see Sec.~\ref{sec:numerical}. Dashed lines are the analytical benchmark, see Sec.~\ref{sec:anlinear}, obtained by collocation methods. Two values of the weight are considered: $M=1$ (blue) and $M=200$ (orange). The solution of the problem with fixed boundary conditions (``full bridge'' case, see Ref.~\cite{SaBaMG2024}) is shown in green. We initialize the neural network as $-\partial U=-q$ and then select the minimal-cost realization out of 2000 epochs. We set $\varepsilon=1$.
    \label{fig:linear_comparison}}
\end{figure}

We first address the case of Gaussian boundary conditions, corresponding to parabolic shapes of $U_i$ and $U_f$. The evolution is linear, and we can compare it to the ground-truth results from Sec.~\ref{sec:anlinear}. These results are shown in Fig.~\ref{fig:linear_comparison} with the solution obtained by the machine learning strategy described in Sec.~\ref{sec:numerical}. We verify (i) that the network is able to reproduce with high accuracy the analytic solution and (ii) that in the limit $M\to \infty$, as expected, the solution converges to that of the full bridge problem.

\begin{figure*}
    \centering
    \includegraphics[width=.9\linewidth]{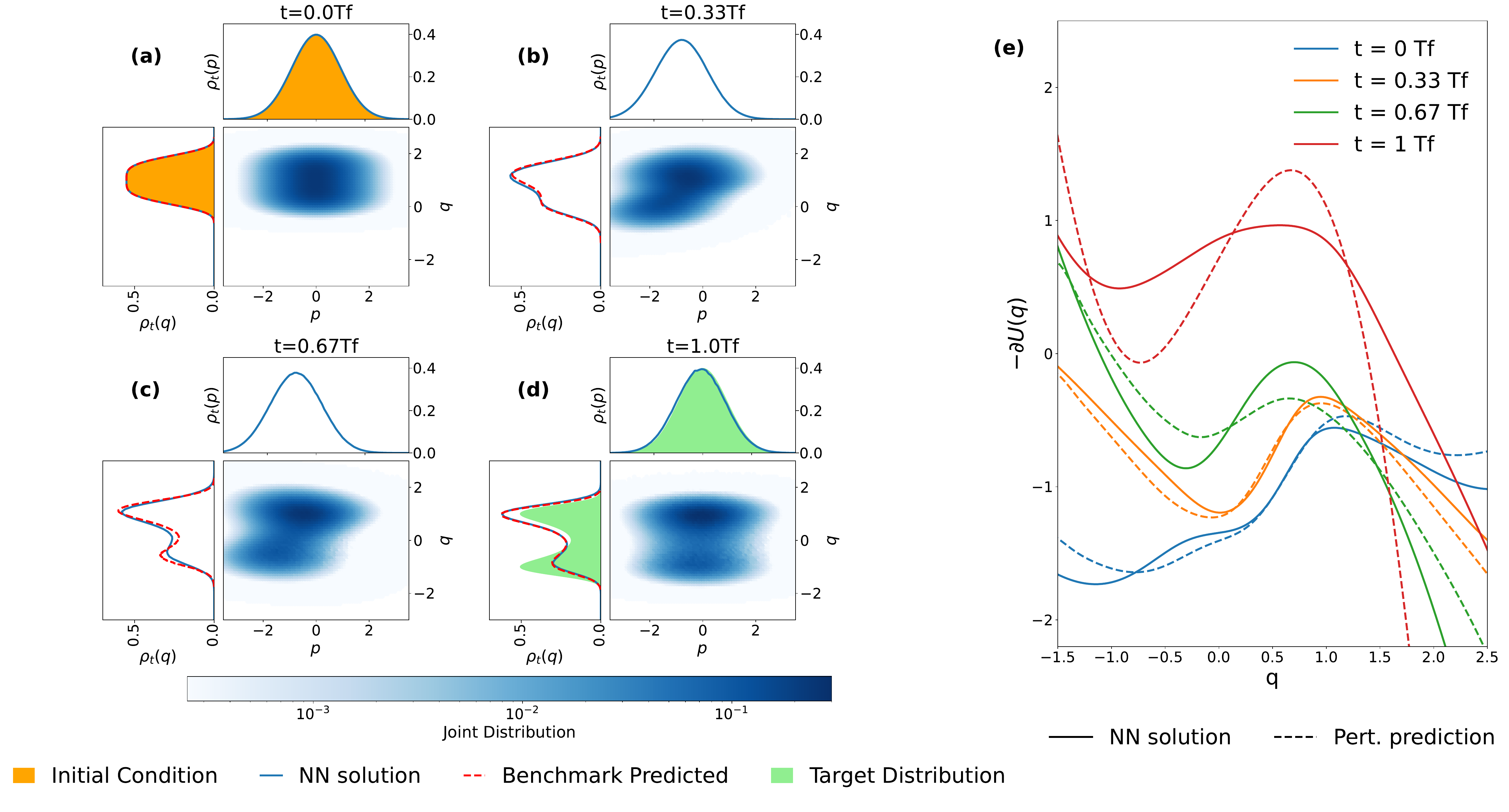}
    \caption{Solution of the optimal problem with non-Gaussian boundary conditions in the perturbative regime $\varepsilon=0.2$, for $M=8$. Panels (a)-(d) show the evolution of the joint and marginal pdf at 4 different times. Neural network solution (blue) and perturbative prediction (red) are compared. Panel (e) shows the drift evolution. Here  $U_i=(q-1)^4$ and $U_f=(q^2-1)^2$. The network is initialized at epoch 0 with the perturbative analytical solution. We train the network for 100 epochs.\label{fig:evo_combo_1}}
    
\end{figure*}

\begin{figure*}[h!]
    \centering
    \includegraphics[width=.9\linewidth]{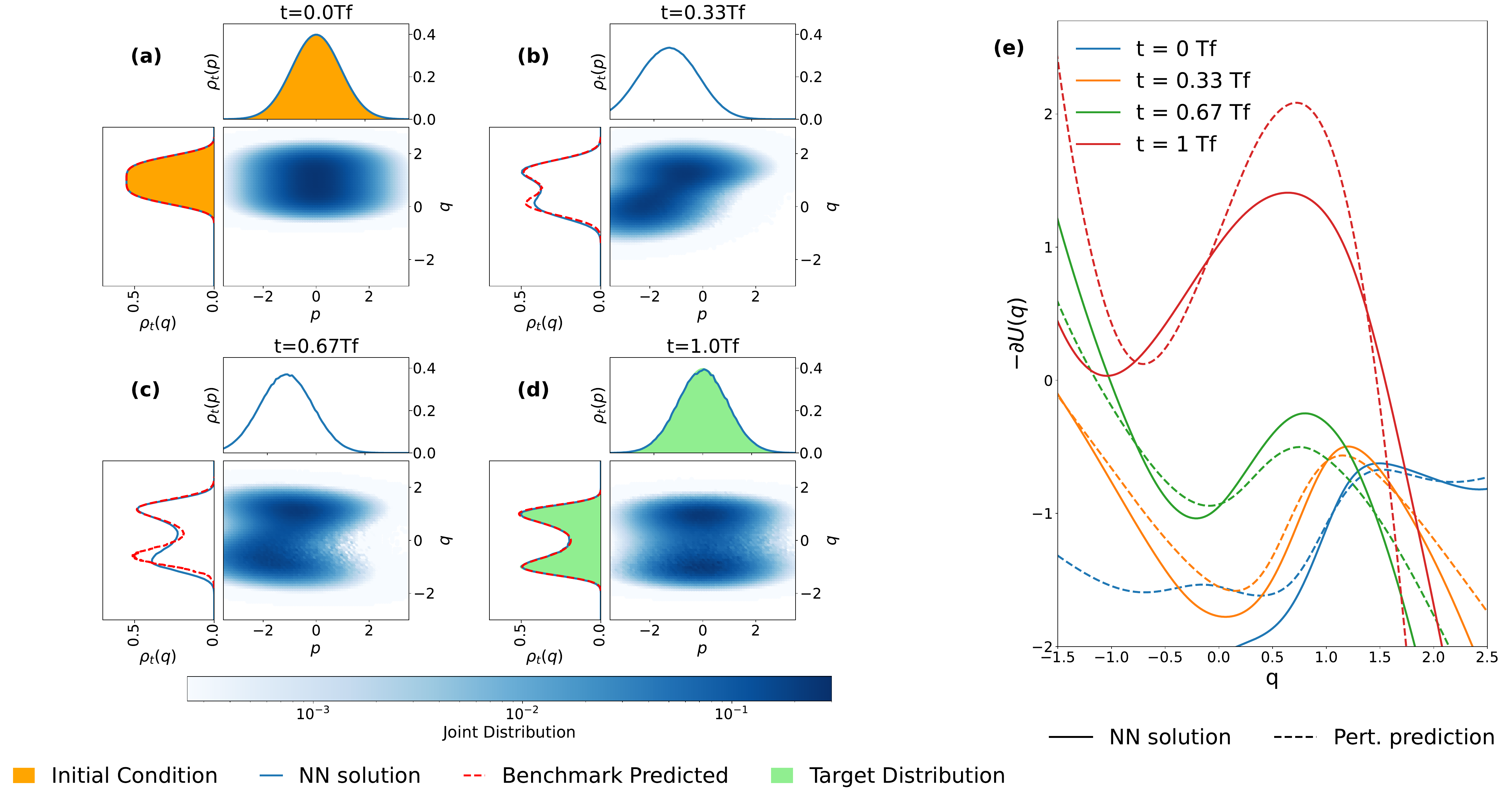}
    \caption{Solution of the optimal problem with non-Gaussian boundary conditions in the  in the perturbative regime, for large $M$. Same as Fig.~\ref{fig:evo_combo_1}, with $M=200$.  The network is initialized with the same drift used in Fig.~\ref{fig:evo_combo_1} (i.e., the analytical solution with $M=8$ and $\varepsilon=0.2$).
    \label{fig:evo_combo_2}}
\end{figure*}

When the boundary conditions are non-Gaussian, we cannot rely on a ground-truth solution, but instead use our multiscale perturbative expansion to approximate the solution described in~\ref{sec:multiscale}. However, for sufficiently large $M$ and small values of $\varepsilon$, we expect that this offers a good approximation. In Figs.~\ref{fig:evo_combo_1} and~\ref{fig:evo_combo_2} we show this comparison for $\varepsilon=0.2$. The agreement between the two methods is good, especially for the position marginal density (leftside plots in panels (a)-(d) in Figs~\ref{fig:evo_combo_1} and~\ref{fig:evo_combo_2}), where the singular expansion provides the most accurate predictions. There are discrepancies in the drift, panels (e) in Figs~\ref{fig:evo_combo_1} and~\ref{fig:evo_combo_2}), especially at later times. The source of this discrepancy is two-fold. Firstly, in our perturbative approach, state-of-the-art multiscale perturbation theory does not provide mathematically rigorous ways to estimate the residuals~\cite{PaSt2008}. Secondly, we expect the non-linear case to be subject to more numerical noise. Because the drift is non-linear, the evolutions of the stochastic differential equation have a larger error from time-discretisation. Evaluation of the final achieved density is less straightforward, requiring more trajectories than in the Gaussian parametric case. As a consequence, we use smaller batch sizes and fewer epochs for training. We intend to examine how this error can be reduced further in future works. 

In Fig.~\ref{fig:evo_combo_2}, the value of $M$ is large enough to expect the solution of the problem to be close to the ``full bridge'' case, where the target distribution is imposed as a boundary condition at the final time. Indeed, the position marginal distribution in panel~\ref{fig:evo_combo_2}(d) is superimposed to the target.

\begin{figure*}
    \centering
    \includegraphics[width=\linewidth]{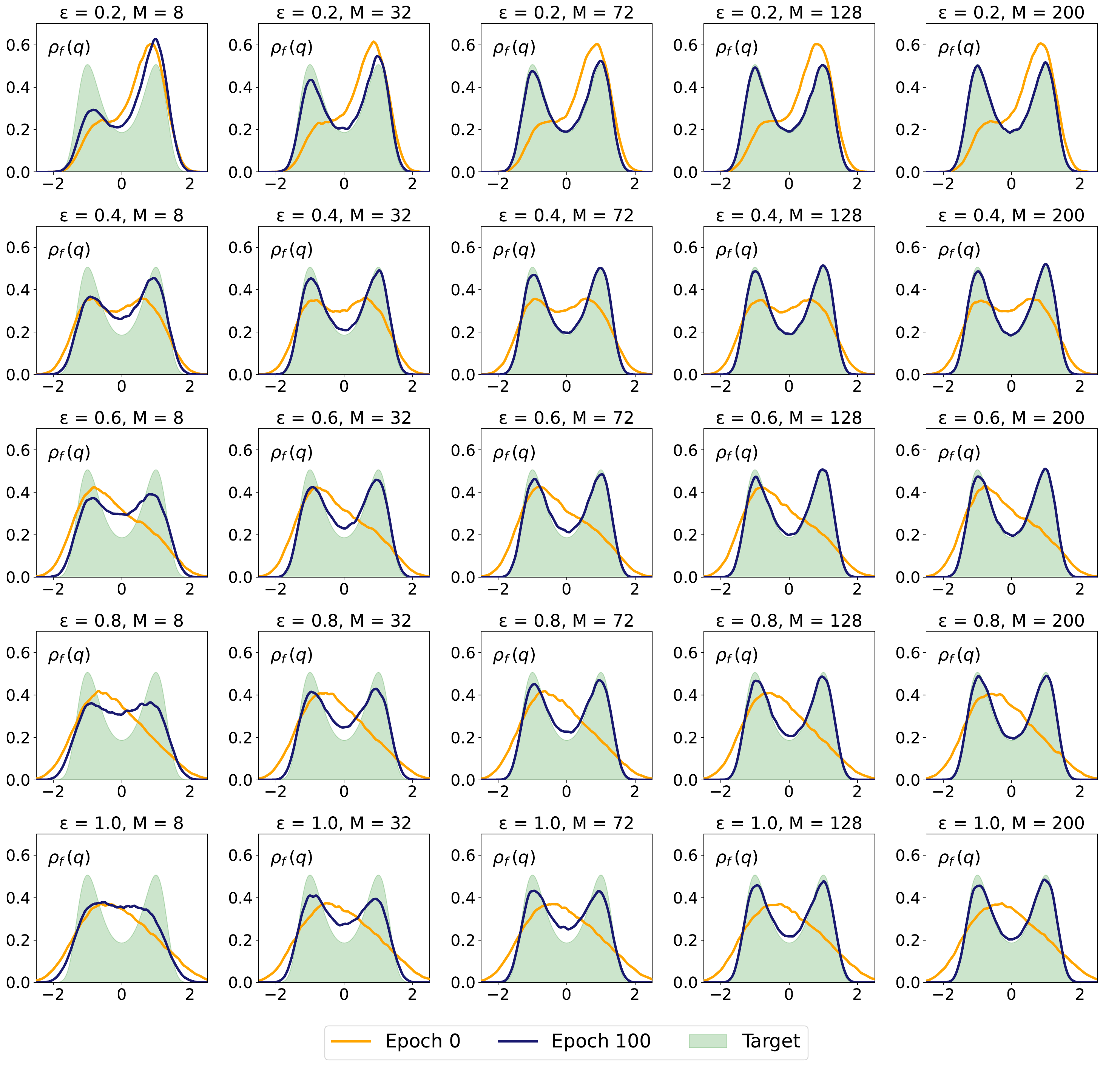}
    \caption{Probability density function at the final time $T$, for different values of $\varepsilon$ and $M$, computed with the machine learning strategy in the presence of non-Gaussian boundary conditions. Blue lines represent the solution at epoch 100, orange lines the solution at epoch 0 (no training). The green shaded area represents the target distribution. Details as in Figs.~\ref{fig:evo_combo_1} and~\ref{fig:evo_combo_2}.
    \label{fig:table_pdf}}
\end{figure*}

\begin{figure*}[h!]
    \centering
    \includegraphics[width=\linewidth]{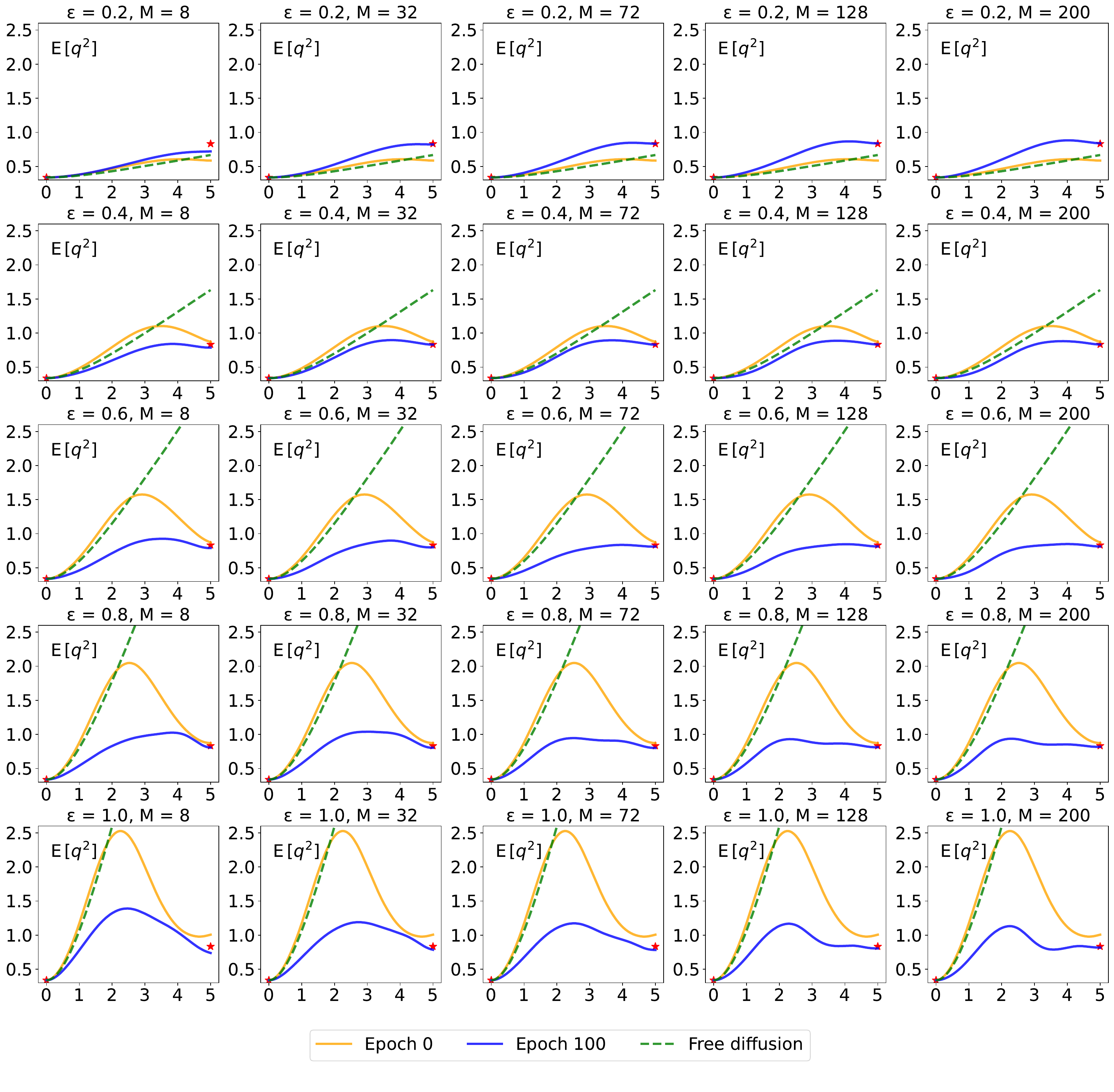}
    \caption{Time-dependent variance, for different values of $\varepsilon$ and $M$, computed with the machine learning strategy in the presence of non-Gaussian boundary conditions. Blue lines represent the solution at epoch 100, orange lines the solution at epoch 0 (no training). The green dashed line represents free diffusion. Details as in Figs.~\ref{fig:evo_combo_1} and~\ref{fig:evo_combo_2}.
    \label{fig:table_var}}
\end{figure*}

Finally, in Figs.~\ref{fig:table_pdf} and~\ref{fig:table_var} we analyze the solution for different values of $\varepsilon$ and $M$. In Fig.~\ref{fig:table_pdf} the position marginal pdf at $T$ is shown. At fixed $\varepsilon$, increasing $M$ allows to get progressively closer to the target distribution at the final time. At fixed $M$, increasing $\varepsilon$ makes the problem stiffer, hence worsening the solution. In Fig.~\ref{fig:table_var} the position variance is shown, as a function of time, for the same cases considered in Fig.~\ref{fig:table_pdf}. It is interesting to see that, with respect to epoch 0 (untrained network), the solution must visibly separate from the free diffusion, hence ``paying'' a larger running cost, in order to match the target distribution and reduce the terminal cost.

\section{Conclusion}

The results presented in this paper for stochastic optimal transport problems in the underdamped dynamics tackle the problem in two distinct but concurrent methods. We approach the problem indirectly, through the first order optimality conditions derived from Pontryagin's maximum principle, and directly, using machine learning to minimize the cost functional. 

Our analytics produce benchmarks for the numerics. When the initial and target distributions are given by Gaussians, the problem is fully characterized by an ordinary differential equation boundary value problem, which we can integrate numerically. In the case of non-harmonic potentials, we use the expansion to find predictions up to second order in $\varepsilon$, that are computed from the numerical solution of a fully integrable cell problem. We want to approximate the full bridge solution by increasing the value of the weighting factor $M$ that penalizes the distance from the target distribution. We demonstrate that this converges to the full bridge for Gaussian boundary conditions. 

Numerically, we can train a neural network to minimise the action functional, using Girsanov's theorem to evaluate this loss without excessive computational overhead, because we do not need to differentiate with respect to simulated trajectories. On comparison with our known solution in for Gaussian boundary conditions, we see good agreement. For non-linear boundary conditions modelling Landauer's bit-erasure problem, we see reasonable agreement with our semi-analytic benchmarks, and are able to reach the target distribution when making $M$ larger.

The numerical method can be applied to different parameters, we can make predictions for more underdamped settings. The training of the neural network is a difficult process. Reducing numerical error in the non-linear training would require a significant increase in computational resources. In addition, we have analytic benchmarks only in the range of parameters where the overdamped expansion are applicable. The analytics and numerics shown in this paper are limited to one dimension, but can be extended to higher dimensions. In future, this methodology could be adapted to different thermodynamical cost functionals. We are in particular interested in the mean entropy production, which requires efficient evaluation of the gradient of a complex neural network with respect to its input. 

We expect that our work will be applicable in multiple fields. Generative diffusion models, in particular, are a very topical issue, and efficiently connecting two data distributions as we do here, may in future also be applicable to lowering energy costs of generative artificial intelligence. Secondly, the methods can be used to inspire efficient protocol design in nanoscale information processing devices.

\begin{acknowledgments}
The authors acknowledge useful discussions with Luca Biferale, Massimo Cencini, Sergio Chibarro, Paolo Andrea Erdman, Marylou Gabri\'e,  Katsiaryna Haitsiukevich, Tuomo Kuusi, Mauro Pastore, Luca Peliti, Kai Puolim\"aki,  and Angelo Vulpiani.
The authors wish to acknowledge CSC – IT Center for Science, Finland, for computational resources.  MB and DL acknowledge the financial support from the MIUR PRIN 2022 (project ``SNO-MINK'' no. 2022KWTEB7) which is funded by the European Union  Next Generation EU, M4 C2 1.1  CUP B53C24006470006. JS acknowledges funding from a University of Helsinki funded doctoral researcher position, Doctoral Programme in Mathematics and Statistic, and partial financial support from the "Finnish Centre of Excellence in Randomness and Structures (First)”.

\emph{Use of artificial intelligence:} We checked our analytic calculations with GPT~6, and used during the development of our codes Claude~Opus~5.5, QWEN~3.8 and Vibe.
\end{acknowledgments}

\section*{References} 
\bibliography{biblio}

\end{document}


\maketitle
\vspace{-10mm}
\begin{abstract}
A weak coupling between position and momentum separates underdamped transport
into rapid momentum relaxation and slow evolution of the position density.
We use this separation to construct an analytical approximation to the
control problem with a Kullback--Leibler terminal cost discussed in~\cite{Main}. The calculation reduces to a cell problem for two functions.
We then obtain the control and phase-space density from time integrals. We derive the control through order $\eps$ and the density
and first- and second-order cumulants through order $\eps^2$. 
\end{abstract}

\section{The control problem and its two time scales}
In contrast to the full-bridge problem of~\cite{Sanders}, here only the initial distribution is prescribed. At the final time, departures from a target distribution are penalized by a terminal cost, so the final distribution may differ from the target. In the dimensionless variables
of the accompanying article~\cite{Main}, the dynamics is
\begin{equation}
 \dd q_t=\eps p_t\dd t,\qquad
 \dd p_t=-[p_t+\eps\partial_qU_t(q_t)]\dd t+\sqrt2\,\dd W_t,
 \qquad 0\leq t\leq T\,,\label{dynamics}
\end{equation}
where the gradient of the potential $\partial_q U_t$ is the control variable. 
With $\phi(p)=(2\pi)^{-1/2}e^{-p^2/2}$, the initial and target densities are
$\rho_i(p,q)=\phi(p)\widetilde\rho_i(q)$ and
$\rho_f(p,q)=\phi(p)\widetilde\rho_f(q)$, where $q,p\in\mathbb R$, $W_t$ is a standard Wiener process,
$\eps,M,T>0$, and $\widetilde\rho_i(q)$, $\widetilde\rho_f(q)$ are normalized densities.
Let $U_\star$ be a given potential defining the reference dynamics, and let $\mathtt f_t$ be the probability density function (pdf) generated by Eq.~\eqref{dynamics} from
$\mathtt f_0=\rho_i$.
The optimal control is defined through minimization of the functional
\begin{equation}
 \mathcal C[U]:=\frac{\eps^2}{4}\int_0^T\!\int_{\mathbb R^2}\mathtt f_t(p,q)\,[\partial_qU_t(q)-\partial_qU_\star(q)]^2\dd p\dd q\dd t
       +M\,\mathrm{KL}(\mathtt f_T\mid\rho_f)\label{cost}
\end{equation}
over admissible potentials for which the dynamics, Eq.~\eqref{dynamics}, and the cost $\mathcal C[U]$ are well defined.
As terminal cost we consider the Kullback--Leibler divergence
\begin{equation}
 \mathrm{KL}(\mathtt f_T\mid\rho_f)
 :=\int_{\mathbb R^2}\mathtt f_T(p,q)
       \log\frac{\mathtt f_T(p,q)}{\rho_f(p,q)}\dd p\dd q.
\end{equation}
Throughout, spatial integrals without limits are over $\mathbb R$.
We assume sufficient smoothness and rapid decay of the densities to justify the calculations. We normalize the value function $V_t$ so that its terminal value is
\begin{equation}
 V_T=M\log(\mathtt f_T/\rho_f),\label{KL}
\end{equation}
which fixes an arbitrary additive constant.

We introduce the fast time $t_0=t$ and the slow time $t_2=\eps^2t$: for $\eps\to0$, momentum relaxes on $t_0$, whereas position changes on $t_2$. Following the full-bridge analysis~\cite{Sanders},
we expand in the probabilists' Hermite polynomials:
\begin{align}
 \mathtt f_t(p,q)&=\phi(p)\sum_{n\geq0}f_t^{(n)}(q)H_n(p),&
 V_t(p,q)&=\sum_{n\geq0}v_t^{(n)}(q)H_n(p),\nonumber\\
 f_t^{(n)}&\sim\sum_{j\geq0}\eps^jf_{t_0,t_2}^{(n:j)},&
 v_t^{(n)}&\sim\sum_{j\geq0}\eps^jv_{t_0,t_2}^{(n:j)},\nonumber\\
 U_t-U_\star&\sim\sum_{j\geq0}\eps^jU_{t_0,t_2}^{(j)}.\label{expansion}
\end{align}
The first index labels the Hermite polynomial and the second the perturbative order in $\eps$. We use the normalization
$H_0=1$, $H_1=p$, $H_2=p^2-1$ and
$\int\phi H_nH_m\dd p=n!\delta_{nm}$~\cite{DLMF}. We also define
$\D g:=\partial_qg+(\partial_qU_\star)g$ and
$\Dt g:=\partial_qg-(\partial_qU_\star)g$ for a generic spatial function $g$. In the expansion, $t_0$ and $t_2$ are treated as independent variables and
$\dd/\dd t=\partial_{t_0}+\eps^2\partial_{t_2}$. The physical solution is
obtained by setting $t_0=t$, $t_2=\eps^2t$. Coefficients independent of the fast time carry their endpoint indices, as in $f_{0,t_2}^{(0:0)}$ and $v_{T,t_2}^{(0:0)}$.
We omit the $q$ dependence where unambiguous.

Before expanding in $\eps$, projection onto the Hermite polynomials gives the optimality equations
\begin{align}
 \partial_tf_t^{(n)}+nf_t^{(n)}
 +\eps[(n+1)\partial_q f_t^{(n+1)}+(\partial_q+\partial_qU_t)f_t^{(n-1)}]&=0,\label{FP}\\
 \partial_tv_t^{(n)}-nv_t^{(n)}
 +\eps[\partial_qv_t^{(n-1)}+(n+1)(\partial_q-\partial_qU_t)v_t^{(n+1)}]
 &=-\delta_{n0}\frac{\eps^2}{4}(\partial_qU_t-\partial_qU_\star)^2,\label{adj}\\
 \frac\eps2(\partial_qU_t-\partial_qU_\star)
 &=\frac{\sum_{n\geq0}(n+1)!f_t^{(n)}v_t^{(n+1)}}{f_t^{(0)}}.\label{stationarity}
\end{align}
Negative Hermite indices are zero. The last equation determines the control.

\section{Boundary conditions near the full bridge}\label{boundary}
When the initial distribution is factorized in $q$ and $p$ and the momentum distribution is Maxwellian, solving the forward hierarchy order by order gives $f_{t_0,t_2}^{(n:j)}=0$ for $j<n$.
Moreover, we use the freedom in the multiscale expansion to set $f_{t_0,t_2}^{(0:1)}=f_{t_0,t_2}^{(1:2)}=0$. 

To second order in $\eps$, the terminal logarithm reads
\begin{align}
 \log\frac{\mathtt f_T}{\rho_f}
 ={}&\log\frac{f_{0,\eps^2T}^{(0:0)}}{\widetilde\rho_f}
 +\eps\frac{f_{T,\eps^2T}^{(1:1)}}{f_{0,\eps^2T}^{(0:0)}}H_1\nonumber\\
 &+\eps^2\left[
 \left(\frac{f_{T,\eps^2T}^{(0:2)}}{f_{0,\eps^2T}^{(0:0)}}-\frac{(f_{T,\eps^2T}^{(1:1)})^2}{2(f_{0,\eps^2T}^{(0:0)})^2}\right)H_0
 +\left(\frac{f_{T,\eps^2T}^{(2:2)}}{f_{0,\eps^2T}^{(0:0)}}-\frac{(f_{T,\eps^2T}^{(1:1)})^2}{2(f_{0,\eps^2T}^{(0:0)})^2}\right)H_2
 \right]+\cdots .\label{log}
\end{align}
Matching the zeroth- and first-order terms yields
\begin{equation}
 v_{T,\eps^2T}^{(0:0)}=M\log\frac{f_{0,\eps^2T}^{(0:0)}}{\widetilde\rho_f},
 \qquad
 v_{T,\eps^2T}^{(1:1)}=M\frac{f_{T,\eps^2T}^{(1:1)}}{f_{0,\eps^2T}^{(0:0)}}.\label{retainedbc}
\end{equation}
The first relation imposes the terminal cost on the position marginal.
The second determines the first-order pdf correction at the final time.
At second order, we choose the boundary condition using the full-bridge limit. In a regular fixed-$M$ expansion, Eq.~\eqref{log} would give
$v_{T,\eps^2T}^{(2:0)}=0$. Here, instead, we keep this coefficient and require the complete second Hermite coefficient 
\begin{equation}
 v_T^{(2)}(q)=\frac M2\int\phi(p)H_2(p)
                 \log\frac{\mathtt f_T(p,q)}{\rho_f(p,q)}\dd p\,,\label{H2exact}
\end{equation}
to have a well-behaved large-$M$ limit. 
At fixed $\eps$, a bounded $v_T^{(2)}$ requires the logarithmic projection
in Eq.~\eqref{H2exact} to be $O(M^{-1})$. 
By considering Eq.~\eqref{log}, this means that at $O(\eps^2)$ the term that survives as $M\to\infty$
\begin{equation}
 \frac{f_{T,\eps^2T}^{(2:2)}}{f_{0,\eps^2T}^{(0:0)}}
       -\frac{(f_{T,\eps^2T}^{(1:1)})^2}{2(f_{0,\eps^2T}^{(0:0)})^2}\label{H2bracket}
\end{equation}
must vanish as $O(M^{-1})$. Since $f_{T,\eps^2T}^{(1:1)}$ vanishes in this limit, the condition reduces to cancelling the first term of Eq.~\eqref{H2bracket} through the choice of $v_{T,t_2}^{(2:0)}$. This is the full-bridge condition on the second-order density correction. Section~\ref{H2selection} derives the
corresponding equation for $v_{T,\eps^2T}^{(2:0)}$ and checks that the finite-$M$ correction leaves the terminal Hermite projection bounded.

The scalar second-order terminal contribution must also be accounted for.
The part supplied by Eq.~\eqref{log} is
\begin{equation}
 v_{T,\eps^2T}^{(0:2)}
 =M\left[\frac{f_{T,\eps^2T}^{(0:2)}}{f_{0,\eps^2T}^{(0:0)}}
       -\frac{(f_{T,\eps^2T}^{(1:1)})^2}{2(f_{0,\eps^2T}^{(0:0)})^2}\right].\label{scalarbc}
\end{equation}
We determine the leading-order functions first, then use this terminal contribution to fix the integration function in the reconstruction of $v^{(0:2)}$. Section~\ref{scalarreconstruction} shows how this agrees with the slow evolution equation and evaluates its large-$M$ limit.

The result is a singular approximation intended for $M\eps^2\gg1$. We derive no error bound uniform in $M$, and the finite-$M$ terms are not a complete expansion of the exact optimum in $M^{-1}$.

\section{Fast relaxation and the leading control}
\subsection{Zeroth order}
At zeroth order, we solve the equations for the pdf and value function:
\begin{equation}
 (\partial_{t_0}+n)f_{t_0,t_2}^{(n:0)}=0,
 \qquad (\partial_{t_0}-n)v_{t_0,t_2}^{(n:0)}=0.
\end{equation}
The Maxwellian initial data and the chosen value-function coefficients give
\begin{align}
 f_{t_0,t_2}^{(0:0)}&=f_{0,t_2}^{(0:0)},&
 f_{t_0,t_2}^{(n:0)}&=0\quad(n\geq1),\nonumber\\
 v_{t_0,t_2}^{(0:0)}&=v_{T,t_2}^{(0:0)},&
 v_{t_0,t_2}^{(2:0)}&=v_{T,t_2}^{(2:0)}e^{-2(T-t_0)}.\label{v20}
\end{align}
Stationarity sets $v_{t_0,t_2}^{(1:0)}=0$. We set the other zeroth-order value-function coefficients to zero in the approximation. At this stage,
$f_{0,t_2}^{(0:0)}$, $v_{T,t_2}^{(0:0)}$ and $v_{T,t_2}^{(2:0)}$ are unknown
functions of position and slow time. We determine them in Section~\ref{cellsection}.

\subsection{First-order equations and endpoint conditions}
We obtain the first-order equations and their boundary conditions from Eqs.~\eqref{FP}--\eqref{stationarity}:
\begin{align}
 \partial_{t_0} f_{t_0,t_2}^{(1:1)}+f_{t_0,t_2}^{(1:1)}+\D f_{0,t_2}^{(0:0)}
 &=-f_{0,t_2}^{(0:0)}\partial_q U_{t_0,t_2}^{(0)},\nonumber\\
 \partial_{t_0} v_{t_0,t_2}^{(1:1)}-v_{t_0,t_2}^{(1:1)}+\partial_q v_{T,t_2}^{(0:0)}+2\Dt v_{t_0,t_2}^{(2:0)}
 &=2(\partial_q U_{t_0,t_2}^{(0)})v_{t_0,t_2}^{(2:0)},\nonumber\\
 \partial_q U_{t_0,t_2}^{(0)}
 &=2v_{t_0,t_2}^{(1:1)}+\frac{4f_{t_0,t_2}^{(1:1)}v_{t_0,t_2}^{(2:0)}}{f_{0,t_2}^{(0:0)}}.\label{orderone}
\end{align}
Substituting the last equation into the first isolates the value-function coefficient:
\begin{equation}
 v_{t_0,t_2}^{(1:1)}=-\frac{\D f_{0,t_2}^{(0:0)}+\partial_{t_0} f_{t_0,t_2}^{(1:1)}
 +(1+4v_{t_0,t_2}^{(2:0)})f_{t_0,t_2}^{(1:1)}}{2f_{0,t_2}^{(0:0)}}.\label{v11}
\end{equation}
We differentiate Eq.~\eqref{v11} with respect to $t_0$, substitute into the value-function equation in Eq.~\eqref{orderone}, and use
$\partial_{t_0}v_{t_0,t_2}^{(2:0)}=2v_{t_0,t_2}^{(2:0)}$. The terms quadratic
in $v_{t_0,t_2}^{(2:0)}$ cancel, leaving
\begin{equation}
 (\partial_{t_0}^2-1)f_{t_0,t_2}^{(1:1)}
 =\D f_{0,t_2}^{(0:0)}+2f_{0,t_2}^{(0:0)}\partial_q v_{T,t_2}^{(0:0)}
 +4e^{-2(T-t_0)}\partial_q(f_{0,t_2}^{(0:0)}v_{T,t_2}^{(2:0)}).\label{fastode}
\end{equation}
The initial first-order pdf correction is zero. Inserting the terminal condition for
$v_{T,t_2}^{(1:1)}$ from Eq.~\eqref{retainedbc} into Eq.~\eqref{v11} gives the second
boundary condition:
\begin{align}
 f_{0,t_2}^{(1:1)}&=0,\nonumber\\
 \left.\partial_{t_0}f_{t_0,t_2}^{(1:1)}\right|_{t_0=T}
 +(2M+1+4v_{T,t_2}^{(2:0)})f_{T,t_2}^{(1:1)}&=-\D f_{0,t_2}^{(0:0)}.\label{robin}
\end{align}
The first-Hermite KL condition determines the terminal pdf correction through this Robin boundary condition.

\subsection{Green-function solution and control}
We obtain the pdf correction and control from the Green function. The Dirichlet inverse of $\partial_{t_0}^2-1$ has kernel
\begin{equation}
 G_{t_0,r}=-\frac{\sinh(\min(t_0,r))\sinh(T-\max(t_0,r))}{\sinh T},
 \qquad 0\leq t_0,r\leq T.
\end{equation}
For the two forcing profiles in Eq.~\eqref{fastode}, define
\begin{align}
 \mathscr G_{t_0}^{(1:1)}&:=\int_0^TG_{t_0,r}\dd r
 =\cosh t_0-1-\sinh t_0\tanh(T/2),\nonumber\\
 \mathscr G_{t_0}^{(1:2)}&:=\int_0^TG_{t_0,r}e^{-2(T-r)}\dd r
 =\frac{e^{-2T}(e^{2t_0}-e^{t_0})}{3}-\frac{2\sinh t_0}{3(1+e^T)}.\label{greens}
\end{align}
The particular solution with zero endpoint values has the fast-time boundary conditions of the full bridge:
\begin{equation}
 \widetilde f_{t_0,t_2}^{(1:1)}
 =[\D f_{0,t_2}^{(0:0)}+2f_{0,t_2}^{(0:0)}\partial_q v_{T,t_2}^{(0:0)}]\mathscr G_{t_0}^{(1:1)}
 +4\partial_q(f_{0,t_2}^{(0:0)}v_{T,t_2}^{(2:0)})\mathscr G_{t_0}^{(1:2)}.\label{tilde}
\end{equation}
Add the homogeneous solution that vanishes at $0$ and equals $X_{M,t_2}$ at $T$:
\begin{equation}
 f_{t_0,t_2}^{(1:1)}=\widetilde f_{t_0,t_2}^{(1:1)}+s(t_0)X_{M,t_2},
 \qquad s(t_0):=\frac{\sinh t_0}{\sinh T}.\label{Jsplit}
\end{equation}
Since $\partial_{t_0}s(T)=\coth T$, Eq.~\eqref{robin} determines
\begin{equation}
 X_{M,t_2}=-\frac{\D f_{0,t_2}^{(0:0)}+
 \left.\partial_{t_0}\widetilde f_{t_0,t_2}^{(1:1)}\right|_{t_0=T}}
 {2M+1+4v_{T,t_2}^{(2:0)}+\coth T}.\label{XM}
\end{equation}
Substituting Eq.~\eqref{tilde} gives the numerator:
\begin{align}
 \D f_{0,t_2}^{(0:0)}+
 \left.\partial_{t_0}\widetilde f_{t_0,t_2}^{(1:1)}\right|_{t_0=T}
 ={}&[1+\tanh(T/2)]\D f_{0,t_2}^{(0:0)}
 +2\tanh(T/2)f_{0,t_2}^{(0:0)}\partial_qv_{T,t_2}^{(0:0)}\nonumber\\
 &+\frac{4(1-e^{-T})(1+2e^{-T})}{3(1+e^{-T})}
       \partial_q(f_{0,t_2}^{(0:0)}v_{T,t_2}^{(2:0)}).\label{numerator}
\end{align}
For bounded fields, the denominator grows as $2M$ and
$X_{M,t_2}=O(M^{-1})$. The solution $\widetilde f_{t_0,t_2}^{(1:1)}$ satisfies the full-bridge
boundary conditions in fast time. The correction $s(t_0)X_{M,t_2}$ vanishes
as $M\to\infty$. At finite $M$, the functions entering this solution still
satisfy the terminal condition in Eq.~\eqref{retainedbc} and may depend on $M$.
We separate the factors of $X_{M,t_2}$ while leaving the functions in the
cell problem unexpanded in $M^{-1}$. At second order, we include the terms
that survive the large-$M$ limit in the slow evolution and keep the
vanishing terms in the fast corrections.

We split the first-order value-function correction in the same way. Writing
$\widetilde v_{t_0,t_2}^{(1:1)}$ for Eq.~\eqref{v11} evaluated with
$\widetilde f_{t_0,t_2}^{(1:1)}$, we obtain
\begin{equation}
 v_{t_0,t_2}^{(1:1)}=\widetilde v_{t_0,t_2}^{(1:1)}
 -\frac{X_{M,t_2}}{2f_{0,t_2}^{(0:0)}}
 \frac{\cosh t_0+(1+4v_{t_0,t_2}^{(2:0)})\sinh t_0}{\sinh T}.\label{v11split}
\end{equation}
The corresponding endpoint correction also vanishes for large $M$.

Finally, the forward equation in Eq.~\eqref{orderone} gives the control directly:
\begin{equation}
 \partial_q U_{t_0,t_2}^{(0)}
 =-\frac{\D f_{0,t_2}^{(0:0)}+\partial_{t_0}f_{t_0,t_2}^{(1:1)}+f_{t_0,t_2}^{(1:1)}}
 {f_{0,t_2}^{(0:0)}}.\label{force}
\end{equation}
Substituting Eqs.~\eqref{tilde}--\eqref{Jsplit} gives
\begin{align}
 \partial_q U_{t_0,t_2}^{(0)}={}&-\partial_q\log f_{0,t_2}^{(0:0)}-\partial_qU_\star\nonumber\\
 &-\bigl[\partial_q\log f_{0,t_2}^{(0:0)}+\partial_qU_\star
                +2\partial_qv_{T,t_2}^{(0:0)}\bigr]
       (\mathscr G_{t_0}^{(1:1)}+\partial_{t_0}\mathscr G_{t_0}^{(1:1)})\nonumber\\
 &-\frac{4\partial_q(f_{0,t_2}^{(0:0)}v_{T,t_2}^{(2:0)})}{f_{0,t_2}^{(0:0)}}
       (\mathscr G_{t_0}^{(1:2)}+\partial_{t_0}\mathscr G_{t_0}^{(1:2)})
 -\frac{X_{M,t_2}}{f_{0,t_2}^{(0:0)}}\frac{e^{t_0}}{\sinh T}.\label{forceX}
\end{align}

\section{Derivation of the slow cell problem}\label{cellsection}
At second order, the two density equations determine the corrections once
the slow fields are known. Their endpoint conditions also select
$v_{T,t_2}^{(2:0)}$ and the slow density equation. The scalar value-function equation then closes the cell problem.

The $H_0$ and $H_2$ projections of Eq.~\eqref{FP} are
\begin{align}
 \partial_{t_0} f_{t_0,t_2}^{(0:2)}
 &=-\partial_{t_2}f_{0,t_2}^{(0:0)}-\partial_qf_{t_0,t_2}^{(1:1)},\nonumber\\
 (\partial_{t_0}+2)f_{t_0,t_2}^{(2:2)}
 &=-(\partial_q+\partial_q U_\star+\partial_q U_{t_0,t_2}^{(0)})f_{t_0,t_2}^{(1:1)}.\label{secondfp}
\end{align}
We extend the Maxwellian initial data to each auxiliary slow-time slice by
setting $f_{0,t_2}^{(1:1)}=f_{0,t_2}^{(0:2)}=f_{0,t_2}^{(2:2)}=0$.
The scalar equation integrates directly. For the second equation, substitution
of Eq.~\eqref{force} gives
\begin{equation}
 (\partial_{t_0}+2)\left[f_{t_0,t_2}^{(2:2)}
       -\frac{(f_{t_0,t_2}^{(1:1)})^2}{2f_{0,t_2}^{(0:0)}}\right]
 =-f_{0,t_2}^{(0:0)}\partial_q
       \left(\frac{f_{t_0,t_2}^{(1:1)}}{f_{0,t_2}^{(0:0)}}\right).
\end{equation}
The two corrections are therefore
\begin{align}
 f_{t_0,t_2}^{(0:2)}&=-t_0\partial_{t_2}f_{0,t_2}^{(0:0)}-\partial_q\int_0^{t_0}f_{r,t_2}^{(1:1)}\dd r,\label{Rbasic}\\
 f_{t_0,t_2}^{(2:2)}&=\frac{(f_{t_0,t_2}^{(1:1)})^2}{2f_{0,t_2}^{(0:0)}}
 -f_{0,t_2}^{(0:0)}\int_0^{t_0} e^{-2(t_0-r)}\partial_q(f_{r,t_2}^{(1:1)}/f_{0,t_2}^{(0:0)})\dd r.\label{Kbasic}
\end{align}
\subsection{Determining the second Hermite coefficient}\label{H2selection}
We determine the second Hermite coefficient from the full-bridge limit, in which the endpoint correction in Eq.~\eqref{Jsplit} vanishes.
The part of the terminal condition from Eq.~\eqref{H2bracket} that survives
this limit requires the contribution to $f^{(2:2)}$ from the solution with
full-bridge boundary conditions to vanish at $T$. We impose the same
cancellation at finite $M$ to determine $v_{T,t_2}^{(2:0)}$. Since
$\widetilde f_{T,t_2}^{(1:1)}=0$, Eq.~\eqref{Kbasic} reduces the rule
to a weighted time integral. The required integrals, together with the
unweighted ones used for the scalar density correction, are
\begin{align}
 \mathscr G_{t_0}^{(1:1:0)}&=\int_0^{t_0}\mathscr G_{r}^{(1:1)}\dd r
 =-t_0+\sinh t_0-(\cosh t_0-1)\tanh(T/2),\nonumber\\
 \mathscr G_{t_0}^{(1:1:4)}&=\int_0^{t_0}\mathscr G_{r}^{(1:2)}\dd r
 =\frac{e^{-2T}(e^{t_0}-1)^2}{6}-\frac{2(\cosh t_0-1)}{3(1+e^T)},\nonumber\\
 \mathscr G_T^{(1:1:1)}&=\int_0^T e^{-2(T-r)}\mathscr G_{r}^{(1:1)}\dd r
 =-\frac{(1-e^{-T})^2}{6}\tanh(T/2),\nonumber\\
 \mathscr G_T^{(1:1:2)}&=\int_0^T e^{-2(T-r)}\mathscr G_{r}^{(1:2)}\dd r
 =-\frac{e^{-T}(2+\cosh T)(1-e^{-T})^2}{18}\tanh(T/2).\label{integrals}
\end{align}
The part of Eq.~\eqref{Kbasic} that survives at $t_0=T$ as $M\to\infty$ gives
\begin{equation}
 0=\partial_q\left[
 \frac{\mathscr G_T^{(1:1:1)}}{f_{0,t_2}^{(0:0)}}
     (\D f_{0,t_2}^{(0:0)}+2f_{0,t_2}^{(0:0)}\partial_qv_{T,t_2}^{(0:0)})
 +\frac{4\mathscr G_T^{(1:1:2)}}{f_{0,t_2}^{(0:0)}}
      \partial_q(f_{0,t_2}^{(0:0)}v_{T,t_2}^{(2:0)})\right].
\end{equation}
The bracket is independent of $q$. Spatial decay determines that integration
constant and yields
\begin{align}
 4\partial_q(f_{0,t_2}^{(0:0)}v_{T,t_2}^{(2:0)})
 &=-\frac{\mathscr G_T^{(1:1:1)}}{\mathscr G_T^{(1:1:2)}}
       [\D f_{0,t_2}^{(0:0)}+2f_{0,t_2}^{(0:0)}\partial_q v_{T,t_2}^{(0:0)}-\varkappa_{t_2} f_{0,t_2}^{(0:0)}],\label{bclosure}\\
 \varkappa_{t_2}&=\int f_{0,t_2}^{(0:0)}[2\partial_q v_{T,t_2}^{(0:0)}+\partial_q U_\star]\dd q.\label{kappadef}
\end{align}
The second equation follows by spatial integration and decay of $f_{0,t_2}^{(0:0)}v_{T,t_2}^{(2:0)}$;
then Eq.~\eqref{bclosure} can be integrated from $-\infty$ to $q$.
This determines $v_{T,t_2}^{(2:0)}$ in terms of the two slow fields. The spatial integration constant
$\varkappa_{t_2}$ enforces compatibility of the spatial boundary conditions.

The relaxation integral for the endpoint correction is
\begin{equation}
 d(T):=\int_0^Te^{-2(T-r)}s(r)\dd r
 =\frac{e^T/6-e^{-T}/2+e^{-2T}/3}{\sinh T}.\label{dterminal}
\end{equation}
Keeping the finite-$M$ correction in Eq.~\eqref{Kbasic} gives
\begin{equation}
 \frac{f_{T,t_2}^{(2:2)}}{f_{0,t_2}^{(0:0)}}
       -\frac{(f_{T,t_2}^{(1:1)})^2}{2(f_{0,t_2}^{(0:0)})^2}
 =-d(T)\partial_q\left(\frac{X_{M,t_2}}{f_{0,t_2}^{(0:0)}}\right).\label{H2retained}
\end{equation}
For $X_{M,t_2}/f_{0,t_2}^{(0:0)}$ and its spatial derivative of order $M^{-1}$,
the contribution to Eq.~\eqref{H2exact} supplied at density order $\eps^2$,
namely $-M\eps^2d(T)\partial_q(X_{M,t_2}/f_{0,t_2}^{(0:0)})$, stays bounded
as $M\to\infty$ at fixed $\eps$. This checks the boundedness required in Section~\ref{boundary} at these
orders. The contribution belongs to the complete Hermite coefficient,
whose expansion includes $v_{T,t_2}^{(2:0)}$.

\subsection{Slow evolution of the density}
We determine the slow density evolution by requiring the part of
$f^{(0:2)}$ that survives as $M\to\infty$ to vanish at both fast-time
endpoints. Using the separation described after Eq.~\eqref{XM}, we apply this requirement to $\widetilde f^{(1:1)}$; the $X_{M,t_2}$ contribution
vanishes in that limit and stays in the fast correction. Evaluation of Eq.~\eqref{Rbasic} at $T$ gives
\begin{equation}
 T\partial_{t_2}f_{0,t_2}^{(0:0)}
 =-\partial_q\int_0^T\widetilde f_{r,t_2}^{(1:1)}\dd r.\label{densitysolvability}
\end{equation}
Substituting Eq.~\eqref{bclosure} into Eq.~\eqref{tilde} gives
\begin{align}
 \widetilde f_{t_0,t_2}^{(1:1)}={}&
 [\D f_{0,t_2}^{(0:0)}+2f_{0,t_2}^{(0:0)}\partial_qv_{T,t_2}^{(0:0)}]
 \left[\mathscr G_{t_0}^{(1:1)}
 -\frac{\mathscr G_T^{(1:1:1)}}{\mathscr G_T^{(1:1:2)}}\mathscr G_{t_0}^{(1:2)}\right]\nonumber\\
 &+\varkappa_{t_2}f_{0,t_2}^{(0:0)}
 \frac{\mathscr G_T^{(1:1:1)}}{\mathscr G_T^{(1:1:2)}}\mathscr G_{t_0}^{(1:2)}.\label{closedcurrent}
\end{align}
Symmetry of $G$ gives $\mathscr G_T^{(1:1:4)}=\mathscr G_T^{(1:1:1)}$.
Thus the two coefficients produced by the average are
\begin{align}
 A&:=-\frac1T\left[\mathscr G_T^{(1:1:0)}
          -\frac{(\mathscr G_T^{(1:1:1)})^2}{\mathscr G_T^{(1:1:2)}}\right]\nonumber\\
 &=1-\frac{3(1+e^{-T})^2}{1+4e^{-T}+e^{-2T}}\frac{\tanh(T/2)}T,\nonumber\\
 B&:=\frac{(\mathscr G_T^{(1:1:1)})^2}{T\mathscr G_T^{(1:1:2)}}=-\frac{\tanh(T/2)}T\frac{(1-e^{-T})^2}{1+4e^{-T}+e^{-2T}}.\label{AB}
\end{align}
Substitution into Eq.~\eqref{densitysolvability} gives
\begin{equation}
 \partial_{t_2}f_{0,t_2}^{(0:0)}-A\partial_q[\D f_{0,t_2}^{(0:0)}+2f_{0,t_2}^{(0:0)}\partial_q v_{T,t_2}^{(0:0)}]
                +B\varkappa_{t_2}\partial_qf_{0,t_2}^{(0:0)}=0.\label{cellF}
\end{equation}
For $T>0$, $0<A<1$ and $A-B=1-2\tanh(T/2)/T>0$.

\subsection{Slow evolution of the scalar value function}
We obtain the slow equation for the value function from the $H_0$ projection of Eq.~\eqref{adj} at second order, using the stationarity condition:
\begin{equation}
 \partial_{t_0}v_{t_0,t_2}^{(0:2)}+\partial_{t_2}v_{T,t_2}^{(0:0)}
 =(v_{t_0,t_2}^{(1:1)})^2-(2f_{t_0,t_2}^{(1:1)} v_{t_0,t_2}^{(2:0)}/f_{0,t_2}^{(0:0)})^2-\Dt v_{t_0,t_2}^{(1:1)}.\label{scalarraw}
\end{equation}
As for the density, only the part that survives the full-bridge limit is
absorbed into the slow equation. We denote the corresponding value-function corrections by a tilde and impose equal fast-endpoint values,
$\widetilde v_{T,t_2}^{(0:2)}=\widetilde v_{0,t_2}^{(0:2)}$.
We allow a nonzero common endpoint function.
Integrating Eq.~\eqref{scalarraw} therefore gives
\begin{equation}
 \partial_{t_2}v_{T,t_2}^{(0:0)}
 =\frac1T\int_0^T\left[
 (\widetilde v_{r,t_2}^{(1:1)})^2
 -\left(\frac{2\widetilde f_{r,t_2}^{(1:1)}v_{r,t_2}^{(2:0)}}{f_{0,t_2}^{(0:0)}}\right)^2
 -\Dt\widetilde v_{r,t_2}^{(1:1)}\right]\dd r.\label{adjointsolvability}
\end{equation}
The value-function correction is defined in Eq.~\eqref{v11split}. To perform
the average, we substitute Eq.~\eqref{v11} with the pdf correction that vanishes
at both endpoints and integrate by parts in fast time:
\begin{align}
 \int_0^T(\partial_{t_0} \widetilde f_{t_0,t_2}^{(1:1)})^2\dd t_0
 &=-\int_0^T\widetilde f_{t_0,t_2}^{(1:1)}\,\partial_{t_0}^2 \widetilde f_{t_0,t_2}^{(1:1)}\dd t_0,\nonumber\\
 \int_0^Tv_{t_0,t_2}^{(2:0)}\widetilde f_{t_0,t_2}^{(1:1)}\,\partial_{t_0} \widetilde f_{t_0,t_2}^{(1:1)}\dd t_0
 &=-\int_0^Tv_{t_0,t_2}^{(2:0)}(\widetilde f_{t_0,t_2}^{(1:1)})^2\dd t_0.\label{averagingibp}
\end{align}
Both identities have zero boundary terms because
$\widetilde f_{0,t_2}^{(1:1)}=\widetilde f_{T,t_2}^{(1:1)}=0$; the second also
uses $\partial_{t_0}v_{t_0,t_2}^{(2:0)}=2v_{t_0,t_2}^{(2:0)}$. We replace
$\partial_{t_0}^2\widetilde f_{t_0,t_2}^{(1:1)}$ by Eq.~\eqref{fastode} in the first identity.
The square of $\widetilde f^{(1:1)}$ cancels, and the remaining quadratic
terms reduce the average to two linear integrals:
\begin{align}
 \partial_{t_2}v_{T,t_2}^{(0:0)}={}&
 \frac{(\D f_{0,t_2}^{(0:0)})^2}{4(f_{0,t_2}^{(0:0)})^2}
 +\frac{\D f_{0,t_2}^{(0:0)}-2f_{0,t_2}^{(0:0)}\partial_qv_{T,t_2}^{(0:0)}}
       {4T(f_{0,t_2}^{(0:0)})^2}\int_0^T\widetilde f_{r,t_2}^{(1:1)}\dd r\nonumber\\
 &+\frac{2v_{T,t_2}^{(2:0)}\D f_{0,t_2}^{(0:0)}
             -\partial_q(f_{0,t_2}^{(0:0)}v_{T,t_2}^{(2:0)})}
       {T(f_{0,t_2}^{(0:0)})^2}
       \int_0^Te^{-2(T-r)}\widetilde f_{r,t_2}^{(1:1)}\dd r\nonumber\\
 &+\frac12\Dt\left[\frac{\D f_{0,t_2}^{(0:0)}}{f_{0,t_2}^{(0:0)}}
       +\frac{1}{Tf_{0,t_2}^{(0:0)}}\int_0^T\widetilde f_{r,t_2}^{(1:1)}\dd r
       +\frac{4v_{T,t_2}^{(2:0)}}{Tf_{0,t_2}^{(0:0)}}
          \int_0^Te^{-2(T-r)}\widetilde f_{r,t_2}^{(1:1)}\dd r\right].\label{scalaraverage}
\end{align}
The remaining integrals are
\begin{align}
 \frac1T\int_0^T\widetilde f_{t_0,t_2}^{(1:1)}\dd t_0
 &=-A[\D f_{0,t_2}^{(0:0)}+2f_{0,t_2}^{(0:0)}\partial_q v_{T,t_2}^{(0:0)}]+B\varkappa_{t_2} f_{0,t_2}^{(0:0)},\nonumber\\
 \int_0^Te^{-2(T-t_0)}\frac{\widetilde f_{t_0,t_2}^{(1:1)}}{f_{0,t_2}^{(0:0)}}\dd t_0
 &=\varkappa_{t_2}\mathscr G_T^{(1:1:1)}.\label{averages}
\end{align}
Together with Eq.~\eqref{bclosure}, these reduce the scalar average to

\begin{align}
 0={}&\partial_{t_2}v_{T,t_2}^{(0:0)}
 +A[\partial_q^2 v_{T,t_2}^{(0:0)}-(\partial_q v_{T,t_2}^{(0:0)})^2-(\partial_q U_\star)(\partial_q v_{T,t_2}^{(0:0)})]\nonumber\\
 &-(1-A)\left[\frac{\partial_q\D f_{0,t_2}^{(0:0)}}{2f_{0,t_2}^{(0:0)}}
                   -\frac{(\D f_{0,t_2}^{(0:0)})^2}{4(f_{0,t_2}^{(0:0)})^2}\right]
 +\frac{B\varkappa_{t_2}}{4}[4\partial_q v_{T,t_2}^{(0:0)}+2\partial_q U_\star-\varkappa_{t_2}].\label{cellV}
\end{align}
Equations~\eqref{cellF} and~\eqref{cellV} are the cell problem.
The terms that vanish in the full-bridge limit remain in the fast corrections. We allow their endpoint values to differ.

\paragraph{Scalar terminal reconstruction.}\label{scalarreconstruction}
We impose the scalar terminal datum in Eq.~\eqref{scalarbc} after determining
the slow equation. This is possible because the initial value of the
value function is unspecified. We keep the terms excluded from the slow
averages because the KL relation multiplies them by $M$.
Integrating Eq.~\eqref{scalarraw} gives
\begin{align}
 v_{t_0,t_2}^{(0:2)}={}&v_{T,t_2}^{(0:2)}
 -\int_{t_0}^T\biggl[(v_{r,t_2}^{(1:1)})^2
 -\left(\frac{2f_{r,t_2}^{(1:1)}v_{r,t_2}^{(2:0)}}{f_{0,t_2}^{(0:0)}}\right)^2
 -\Dt v_{r,t_2}^{(1:1)}-\partial_{t_2}v_{T,t_2}^{(0:0)}\biggr]\dd r.\label{scalarreconstructed}
\end{align}
From Eqs.~\eqref{Rbasic} and~\eqref{densitysolvability},
$f_{T,t_2}^{(0:2)}=-\tanh(T/2)\partial_qX_{M,t_2}$ and
$f_{T,t_2}^{(1:1)}=X_{M,t_2}$. Thus Eq.~\eqref{scalarbc} gives
\begin{equation}
 v_{T,t_2}^{(0:2)}=-M\tanh(T/2)\frac{\partial_qX_{M,t_2}}{f_{0,t_2}^{(0:0)}}
             -\frac{MX_{M,t_2}^2}{2(f_{0,t_2}^{(0:0)})^2}.\label{scalarexplicit}
\end{equation}
For functions in the cell problem with a regular large-$M$ limit,
assume that the ratios in Eq.~\eqref{XM} and their spatial derivatives remain
bounded. Expanding Eq.~\eqref{scalarexplicit} gives
\begin{equation}
 v_{T,t_2}^{(0:2)}
 =\frac{\tanh(T/2)}{2f_{0,t_2}^{(0:0)}}\partial_q\left[
 \D f_{0,t_2}^{(0:0)}
 +\left.\partial_{t_0}\widetilde f_{t_0,t_2}^{(1:1)}\right|_{t_0=T}\right]
 +O(M^{-1}).\label{scalarlimit}
\end{equation}
This terminal value is finite but generally nonzero. Keeping the scalar
correction accounts for a discrepancy that can persist as $M\to\infty$.
Its value fixes a spatially dependent integration function in
Eq.~\eqref{scalarreconstructed}, separately from the boundary conditions for
the leading-order functions and the additive constant in $v_{T,t_2}^{(0:0)}$. Its spatial derivative first enters the first-Hermite equation for the value function at order $\eps^3$, so restoring this terminal datum leaves the control and density unchanged at these orders.
Under the same boundedness assumptions, the right-hand side of
Eq.~\eqref{scalarraw} changes by $O(M^{-1})$ when we replace the first-order
coefficients by their tilde parts. Consequently,
\begin{equation}
 v_{T,t_2}^{(0:2)}-v_{0,t_2}^{(0:2)}=O(M^{-1}),\label{scalarendpoints}
\end{equation}
which recovers the equal-endpoint condition in the full-bridge limit while keeping the scalar terminal contribution at finite $M$.

\subsection{Solving the cell problem}
We rewrite the cell problem by absorbing the density-gradient terms in Eq.~\eqref{cellV} into an auxiliary variable. Set
\begin{equation}
 v_{T,t_2}^{(0:0)}=\frac{\sigma_{t_2}}{2A}-\frac{U_\star}{2}
 +\frac{B\varkappa_{t_2} q}{2A}
 +\alpha\log f_{0,t_2}^{(0:0)}+\jmath_{t_2},\qquad
 \alpha=\frac{1-\sqrt A}{2\sqrt A}.\label{transform}
\end{equation}
Substitution into Eqs.~\eqref{cellF} and~\eqref{cellV}, using
$A(1+2\alpha)=\sqrt A$ and $A+4A\alpha(1+\alpha)=1$, yields
\begin{align}
 \partial_{t_2}f_{0,t_2}^{(0:0)}-\partial_q(f_{0,t_2}^{(0:0)}\partial_q \sigma_{t_2})-\sqrt A\,\partial_q^2f_{0,t_2}^{(0:0)}&=0,\label{sigmaF}\\
 \partial_{t_2}\sigma_{t_2}-\tfrac12(\partial_q \sigma_{t_2})^2+\sqrt A\,\partial_q^2 \sigma_{t_2}
       +\tfrac A2(\partial_q U_\star)^2-A\partial_q^2 U_\star+Bq\partial_{t_2}\varkappa_{t_2}&=0,\label{sigmaV}\\
 (A-B)\varkappa_{t_2}=\int f_{0,t_2}^{(0:0)}\partial_q \sigma_{t_2}\dd q,
 \qquad \partial_{t_2}\jmath_{t_2}&=\frac{B(A-B)}{4A}\varkappa_{t_2}^2.\label{gauge}
\end{align}
Differentiating the first relation in Eq.~\eqref{gauge} and integrating by parts
in Eqs.~\eqref{sigmaF} and~\eqref{sigmaV} gives
\begin{equation}
 \partial_{t_2}\varkappa_{t_2}=\int f_{0,t_2}^{(0:0)}[-(\partial_q U_\star)(\partial_q^2 U_\star)+\partial_q^3 U_\star]\dd q.\label{kdot}
\end{equation}
The initial datum is $f_{0,0}^{(0:0)}(q)=\widetilde\rho_i(q)$.
The scalar KL boundary condition at the final slow time becomes
\begin{equation}
 \sigma_{\eps^2T}=2A(M-\alpha)\log f_{0,\eps^2T}^{(0:0)}-2AM\log\widetilde\rho_f
           +AU_\star-B\varkappa_{\eps^2T}q-2A\jmath_{\eps^2T}.\label{sigmabc}
\end{equation}
We fix $\jmath_0=0$ in Eq.~\eqref{transform}; its evolution in Eq.~\eqref{gauge}
then determines the scalar term above. The term linear in $q$ accounts
for motion of the mean position.

The Cole--Hopf variables
$h_{t_2}=\exp[-\sigma_{t_2}/(2\sqrt A)]$, $\widehat h_{t_2}=f_{0,t_2}^{(0:0)}/h_{t_2}$ obey
\begin{align}
 \partial_{t_2}h_{t_2}&=-\sqrt A\,\partial_q^2 h_{t_2}+
 \frac{\frac A2(\partial_q U_\star)^2-A\partial_q^2 U_\star+Bq\partial_{t_2}\varkappa_{t_2}}{2\sqrt A}\,h_{t_2},\nonumber\\
 \partial_{t_2}\widehat h_{t_2}&=\sqrt A\,\partial_q^2 \widehat h_{t_2}-
 \frac{\frac A2(\partial_q U_\star)^2-A\partial_q^2 U_\star+Bq\partial_{t_2}\varkappa_{t_2}}{2\sqrt A}\,\widehat h_{t_2}.\label{colehopf}
\end{align}
For $U_\star=0$, Eq.~\eqref{kdot} makes $\varkappa_{t_2}$ constant and these are
backward/forward heat equations coupled through their boundary data.
For general $U_\star$, $\varkappa_{t_2}$ can vary with time and depends
on the density through Eq.~\eqref{kdot}. This dependence couples the two equations.

\section{Phase-space density and position marginal}
The solution of the cell problem gives $f_{0,t_2}^{(0:0)}$ and
$v_{T,t_2}^{(0:0)}$. Equation~\eqref{bclosure} then determines
$v_{T,t_2}^{(2:0)}$, and Eqs.~\eqref{tilde}--\eqref{XM} determine the first-order pdf correction.
The density follows from Eqs.~\eqref{Rbasic} and~\eqref{Kbasic}.
The time integrals of the endpoint-correction profile are
\begin{align}
 c(t_0)&:=\int_0^{t_0}s(r)\dd r=\frac{\cosh t_0-1}{\sinh T},\nonumber\\
 d(t_0)&:=\int_0^{t_0}e^{-2(t_0-r)}s(r)\dd r
 =\frac{e^{t_0}/6-e^{-t_0}/2+e^{-2t_0}/3}{\sinh T}.\label{sd}
\end{align}
Using the slow density equation in Eq.~\eqref{Rbasic} gives
\begin{align}
 f_{t_0,t_2}^{(0:2)}={}&-\partial_q[\D f_{0,t_2}^{(0:0)}+2f_{0,t_2}^{(0:0)}\partial_q v_{T,t_2}^{(0:0)}]
 \left[\mathscr G_{t_0}^{(1:1:0)}-\frac{t_0}{T}\mathscr G_T^{(1:1:0)}\right]\nonumber\\
 &-4\partial_q^2(f_{0,t_2}^{(0:0)}v_{T,t_2}^{(2:0)})
 \left[\mathscr G_{t_0}^{(1:1:4)}-\frac{t_0}{T}\mathscr G_T^{(1:1:4)}\right]
 -c(t_0)\partial_qX_{M,t_2}.\label{Rexplicit}
\end{align}
For $f^{(2:2)}$, all the time integrations in Eq.~\eqref{Kbasic} can likewise be
performed. Extend the weighted Green integrals in Eq.~\eqref{integrals} to
$0\leq t_0\leq T$:
\begin{align}
 \mathscr G_{t_0}^{(1:1:1)}&:=\int_0^{t_0}e^{-2(t_0-r)}\mathscr G_r^{(1:1)}\dd r\nonumber\\
 &=\frac{e^{t_0}-e^{-2t_0}}{3(1+e^T)}
   +\frac{e^T(e^{-t_0}-e^{-2t_0})}{1+e^T}
   -\frac{1-e^{-2t_0}}2,\nonumber\\
 \mathscr G_{t_0}^{(1:1:2)}&:=\int_0^{t_0}e^{-2(t_0-r)}\mathscr G_r^{(1:2)}\dd r\nonumber\\
 &=e^{-2T}\left[\frac{e^{2t_0}-e^{-2t_0}}{12}
                  -\frac{e^{t_0}-e^{-2t_0}}9\right]
   -\frac{e^{t_0}-3e^{-t_0}+2e^{-2t_0}}{9(1+e^T)}.\label{weightedprofiles}
\end{align}
Substitution of Eq.~\eqref{closedcurrent} into Eq.~\eqref{Kbasic} yields
\begin{align}
 f_{t_0,t_2}^{(2:2)}={}&\frac{(f_{t_0,t_2}^{(1:1)})^2}{2f_{0,t_2}^{(0:0)}}
 -f_{0,t_2}^{(0:0)}\partial_q
 \left(\frac{\D f_{0,t_2}^{(0:0)}}{f_{0,t_2}^{(0:0)}}+2\partial_qv_{T,t_2}^{(0:0)}\right)
 \left[\mathscr G_{t_0}^{(1:1:1)}
 -\frac{\mathscr G_T^{(1:1:1)}}{\mathscr G_T^{(1:1:2)}}\mathscr G_{t_0}^{(1:1:2)}\right]\nonumber\\
 &-f_{0,t_2}^{(0:0)}d(t_0)\partial_q
             \left(\frac{X_{M,t_2}}{f_{0,t_2}^{(0:0)}}\right).\label{Kexplicit}
\end{align}
Thus the reconstructed phase-space density and its position marginal are
\begin{align}
 \mathtt f_t(p,q)&\simeq
 \left.\phi(p)\bigl[f_{0,t_2}^{(0:0)}(q)+\eps f_{t_0,t_2}^{(1:1)}(q)H_1(p)
 +\eps^2\bigl(f_{t_0,t_2}^{(0:2)}(q)+f_{t_0,t_2}^{(2:2)}(q)H_2(p)\bigr)\bigr]
 \right|_{t_0=t,\,t_2=\eps^2t},\label{density}\\
 \int\mathtt f_t(p,q)\dd p&\simeq
 \left[f_{0,t_2}^{(0:0)}(q)+\eps^2 f_{t_0,t_2}^{(0:2)}(q)\right]_{t_0=t,\,t_2=\eps^2t}.\label{positiondensity}
\end{align}
Here $\simeq$ denotes the approximation through these orders. The
approximation preserves total mass because Eq.~\eqref{Rexplicit} is a spatial
derivative, so $\int f_{t_0,t_2}^{(0:2)}\dd q=0$. The Hermite truncation allows negative values in arbitrarily remote momentum tails.

At the final fast time, the two corrections simplify to
\begin{equation}
 f_{T,t_2}^{(0:2)}=-c(T)\partial_qX_{M,t_2},\qquad
 f_{T,t_2}^{(2:2)}=\frac{X_{M,t_2}^2}{2f_{0,t_2}^{(0:0)}}
 -f_{0,t_2}^{(0:0)}d(T)\partial_q\left(\frac{X_{M,t_2}}{f_{0,t_2}^{(0:0)}}\right).\label{terminaldensity}
\end{equation}
They vanish as $X_{M,t_2}$ and its derivatives tend to zero. If the scalar value-function coefficient remains bounded as $M\to\infty$, Eq.~\eqref{retainedbc} also gives
$f_{0,\eps^2T}^{(0:0)}\to\widetilde\rho_f$. The reconstructed endpoint then
approaches the full bridge through these orders in the density.

\section{Cumulants}
Integrating the Hermite expansion over momentum gives
\begin{align}
 \int p\mathtt f_t(p,q)\dd p&\simeq\eps f_{t_0,t_2}^{(1:1)},\nonumber\\
 \int p^2\mathtt f_t(p,q)\dd p&\simeq
 f_{0,t_2}^{(0:0)}+\eps^2(f_{t_0,t_2}^{(0:2)}+2f_{t_0,t_2}^{(2:2)}).\label{rawmom}
\end{align}
In this section every right-hand side is evaluated at $t_0=t$,
$t_2=\eps^2t$, after performing the fast-time integrals at fixed $t_2$.
Define the slow position moments and half the rate of change of their variance:
\begin{equation}
 \mu_{t_2}^{(n)}:=\int q^n f_{0,t_2}^{(0:0)}(q)\dd q,\qquad
 \mathcal K_{t_2}:=\frac12\partial_{t_2}
       [\mu_{t_2}^{(2)}-(\mu_{t_2}^{(1)})^2].\label{momdefs}
\end{equation}
Spatial integration of the cell equation gives
\begin{align}
 \partial_{t_2}\mu_{t_2}^{(1)}&=-(A-B)\varkappa_{t_2},\nonumber\\
 \mathcal K_{t_2}&=\sqrt A-\int(q-\mu_{t_2}^{(1)})
                       f_{0,t_2}^{(0:0)}\partial_q\sigma_{t_2}\dd q.\label{slowmoments}
\end{align}
For the remaining spatial integrals, Eqs.~\eqref{closedcurrent} and~\eqref{transform} imply
\begin{align}
 \int\widetilde f_{t_0,t_2}^{(1:1)}\dd q
 &=\varkappa_{t_2}\mathscr G_{t_0}^{(1:1)},\nonumber\\
 \int(q-\mu_{t_2}^{(1)})\widetilde f_{t_0,t_2}^{(1:1)}\dd q
 &=-\frac{\mathcal K_{t_2}}{A}
 \left[\mathscr G_{t_0}^{(1:1)}
       -\frac{\mathscr G_T^{(1:1:1)}}{\mathscr G_T^{(1:1:2)}}\mathscr G_{t_0}^{(1:2)}\right].\label{currentmom}
\end{align}
For example, the centered spatial integral of
$\D f_{0,t_2}^{(0:0)}+2f_{0,t_2}^{(0:0)}\partial_qv_{T,t_2}^{(0:0)}$
is $-\mathcal K_{t_2}/A$; the corresponding integral of
$4\partial_q(f_{0,t_2}^{(0:0)}v_{T,t_2}^{(2:0)})$ is
$\mathcal K_{t_2}\mathscr G_T^{(1:1:1)}/(A\mathscr G_T^{(1:1:2)})$.
These two integrations by parts determine the signs in the covariance and
position variance below.

When the expansion of Eq.~\eqref{XM} can be integrated term by term, the two
moments of the endpoint correction are, to leading order in $M^{-1}$,
\begin{align}
 \int X_{M,t_2}\dd q
 &=-\frac{1}{2M}\left[
       \int f_{0,t_2}^{(0:0)}\partial_qU_\star\dd q
       +\varkappa_{t_2}\tanh(T/2)\right]+O(M^{-2}),\nonumber\\
 \int(q-\mu_{t_2}^{(1)})X_{M,t_2}\dd q
 &=\frac{1}{2M}\biggl[1-\int(q-\mu_{t_2}^{(1)})f_{0,t_2}^{(0:0)}\partial_qU_\star\dd q\nonumber\\
 &\hspace{14mm}+\frac{\mathcal K_{t_2}}{A}
 \left(\tanh(T/2)-\frac{\mathscr G_T^{(1:1:1)}}{\mathscr G_T^{(1:1:2)}}
       \frac{(1-e^{-T})(1+2e^{-T})}{3(1+e^{-T})}\right)\biggr]+O(M^{-2}).\label{Xmoments}
\end{align}
These expansions hold at fixed functions in the cell problem. In the cumulants, we keep $X_{M,t_2}$ with its full denominator from Eq.~\eqref{XM}.

Using Eqs.~\eqref{Jsplit} and~\eqref{currentmom} in the momentum integral gives
\begin{equation}
 \E p_t\simeq\eps\left[
 \varkappa_{t_2}\mathscr G_{t_0}^{(1:1)}+s(t_0)\int X_{M,t_2}\dd q\right].\label{meanp}
\end{equation}
Integrating $qf^{(0:2)}$ from Eq.~\eqref{Rexplicit} once by parts gives the mean position:
\begin{equation}
 \E q_t\simeq\mu_{t_2}^{(1)}
 +\eps^2\varkappa_{t_2}\left[\mathscr G_{t_0}^{(1:1:0)}
                 -\frac{t_0}{T}\mathscr G_T^{(1:1:0)}\right]
 +\eps^2c(t_0)\int X_{M,t_2}\dd q.\label{meanq}
\end{equation}
For the position variance, integrate $(q-\mu_{t_2}^{(1)})^2f^{(0:2)}$;
for the covariance, integrate $\eps(q-\mu_{t_2}^{(1)})f^{(1:1)}$. The results are
\begin{align}
 \Var q_t\simeq{}&\mu_{t_2}^{(2)}-(\mu_{t_2}^{(1)})^2\nonumber\\
 &-\frac{2\eps^2\mathcal K_{t_2}}{A}
 \left[\mathscr G_{t_0}^{(1:1:0)}-\frac{t_0}{T}\mathscr G_T^{(1:1:0)}
 -\frac{\mathscr G_T^{(1:1:1)}}{\mathscr G_T^{(1:1:2)}}
       \left(\mathscr G_{t_0}^{(1:1:4)}-\frac{t_0}{T}\mathscr G_T^{(1:1:4)}\right)\right]\nonumber\\
 &+2\eps^2c(t_0)\int(q-\mu_{t_2}^{(1)})X_{M,t_2}\dd q,\label{varq}\\
 \Cov(q_t,p_t)\simeq{}&-\frac{\eps\mathcal K_{t_2}}{A}
 \left[\mathscr G_{t_0}^{(1:1)}
       -\frac{\mathscr G_T^{(1:1:1)}}{\mathscr G_T^{(1:1:2)}}\mathscr G_{t_0}^{(1:2)}\right]
 +\eps s(t_0)\int(q-\mu_{t_2}^{(1)})X_{M,t_2}\dd q.\label{covqp}
\end{align}
Finally, the second momentum moment and the zero mass of $f^{(0:2)}$ give
\begin{equation}
 \Var p_t\simeq1+\eps^2\left[
 2\int f_{t_0,t_2}^{(2:2)}\dd q
 -\left(\int f_{t_0,t_2}^{(1:1)}\dd q\right)^2\right].\label{varp}
\end{equation}
To evaluate the square and subtract the mean, we first center the first-order pdf correction. From Eqs.~\eqref{closedcurrent} and~\eqref{transform},
\begin{align}
 \frac{f_{t_0,t_2}^{(1:1)}}{f_{0,t_2}^{(0:0)}}={}&
 \varkappa_{t_2}\mathscr G_{t_0}^{(1:1)}
 +\left[\mathscr G_{t_0}^{(1:1)}
       -\frac{\mathscr G_T^{(1:1:1)}}{\mathscr G_T^{(1:1:2)}}\mathscr G_{t_0}^{(1:2)}\right]
 \left[\frac{\partial_q\log f_{0,t_2}^{(0:0)}}{\sqrt A}
       +\frac{\partial_q\sigma_{t_2}-(A-B)\varkappa_{t_2}}A\right]\nonumber\\
 &+s(t_0)\frac{X_{M,t_2}}{f_{0,t_2}^{(0:0)}}.\label{centeredcurrent}
\end{align}
The spatial average of the second bracket with density $f_{0,t_2}^{(0:0)}$
is zero by Eq.~\eqref{gauge}. Substitution into Eqs.~\eqref{Kexplicit} and~\eqref{varp} now gives the final expression
\begin{align}
 \Var p_t\simeq{}&1
 +\eps^2\left[\mathscr G_{t_0}^{(1:1)}
       -\frac{\mathscr G_T^{(1:1:1)}}{\mathscr G_T^{(1:1:2)}}\mathscr G_{t_0}^{(1:2)}\right]^2
 \int f_{0,t_2}^{(0:0)}
 \left[\frac{\partial_q\log f_{0,t_2}^{(0:0)}}{\sqrt A}
       +\frac{\partial_q\sigma_{t_2}-(A-B)\varkappa_{t_2}}A\right]^2\dd q\nonumber\\
 &-2\eps^2\left[\mathscr G_{t_0}^{(1:1:1)}
       -\frac{\mathscr G_T^{(1:1:1)}}{\mathscr G_T^{(1:1:2)}}\mathscr G_{t_0}^{(1:1:2)}\right]
 \int f_{0,t_2}^{(0:0)}\partial_q^2
       \left[\frac{\log f_{0,t_2}^{(0:0)}}{\sqrt A}+\frac{\sigma_{t_2}}A\right]\dd q\nonumber\\
 &+2\eps^2s(t_0)\left[\mathscr G_{t_0}^{(1:1)}
       -\frac{\mathscr G_T^{(1:1:1)}}{\mathscr G_T^{(1:1:2)}}\mathscr G_{t_0}^{(1:2)}\right]
 \int\left[\frac{\partial_q\log f_{0,t_2}^{(0:0)}}{\sqrt A}
       +\frac{\partial_q\sigma_{t_2}-(A-B)\varkappa_{t_2}}A\right]X_{M,t_2}\dd q\nonumber\\
 &-2\eps^2d(t_0)\int f_{0,t_2}^{(0:0)}\partial_q
            \left(\frac{X_{M,t_2}}{f_{0,t_2}^{(0:0)}}\right)\dd q\nonumber\\
 &+\eps^2s(t_0)^2\left[\int\frac{X_{M,t_2}^2}{f_{0,t_2}^{(0:0)}}\dd q
                 -\left(\int X_{M,t_2}\dd q\right)^2\right].\label{varpfull}
\end{align}
The first two correction terms give the result with zero endpoint correction.
The next two are, respectively, the centered cross term and the response of
the relaxation integral; both are $O(\eps^2/M)$ for bounded fields. The last
term is $O(\eps^2/M^2)$. The mixed terms can remain when the mean momentum is zero.

The fast derivatives of the unweighted Green integrals are the profiles
$\mathscr G^{(1:1)}$ and $\mathscr G^{(1:2)}$. Together with
Eq.~\eqref{slowmoments}, this verifies the kinematic relations
\begin{equation}
 \frac{\dd}{\dd t}\E q_t=\eps\E p_t,\qquad
 \frac{\dd}{\dd t}\Var q_t=2\eps\Cov(q_t,p_t)\label{kinematic}
\end{equation}
through these orders, using
$\dd/\dd t=\partial_{t_0}+\eps^2\partial_{t_2}$ at physical time.

At $t=T$, both the unweighted and weighted Green combinations in
Eq.~\eqref{varpfull} vanish. In particular,
\begin{equation}
 \Var p_T\simeq1+\eps^2\left[
 \int\frac{X_{M,\eps^2T}^2}{f_{0,\eps^2T}^{(0:0)}}\dd q
 -\left(\int X_{M,\eps^2T}\dd q\right)^2
 -2d(T)\int f_{0,\eps^2T}^{(0:0)}\partial_q
       \left(\frac{X_{M,\eps^2T}}{f_{0,\eps^2T}^{(0:0)}}\right)\dd q\right].\label{varpterminal}
\end{equation}
The momentum-variance correction also vanishes as
$X_{M,\eps^2T}$ and its spatial derivatives tend to zero in the full-bridge
limit.

\section*{Use of AI}
We checked our analytic calculations with GPT~6, and used LLMs to edit this draft. The authors take full responsibility for the final text.